\documentclass[12pt]{iopart}
\usepackage{iopams}
\usepackage{graphicx}
\usepackage{subfigure}

\begin{document}

\title[QPOs in brany neutron stars]
{Quasiperiodic oscillations in a strong gravitational field around
neutron stars testing braneworld models}

\author{Andrea Kotrlov\'{a}, Zden\v{e}k Stuchl\'{\i}k and Gabriel T\"{o}r\"{o}k}

\address{Institute of
Physics, Faculty of Philosophy and Science, Silesian University in
Opava, Bezru\v{c}ovo n\'am.~13, CZ-746\,01 Opava, Czech Republic}
\ead{Andrea.Kotrlova@fpf.slu.cz}

\begin{abstract}
The strong gravitational field of neutron stars in the brany
universe could be described by spherically symmetric solutions
with a metric in the exterior to the brany stars being of the
Reissner--Nordstr\"{o}m type containing a brany tidal charge
representing the tidal effect of the bulk spacetime onto the star
structure. We investigate the role of the tidal charge in orbital
models of high-frequency quasiperiodic oscillations (QPOs)
observed in neutron star binary systems. We focus on the
relativistic precession model. We give the radial profiles of
frequencies of the Keplerian (vertical) and radial epicyclic
oscillations. We show how the standard relativistic precession
model modified by the tidal charge fits the observational data,
giving estimates of the allowed values of the tidal charge and the
brane tension based on the processes going in the vicinity of
neutron stars. We compare the strong field regime restrictions
with those given in the weak-field limit of solar system
experiments.
\end{abstract}


\maketitle

\section{Introduction}

The string and M-theories describing gravity as higher-dimensional
interaction appearing to be effectively 4D at low energies
inspired braneworld models with the observable universe being a
3-brane (domain wall) to which the standard non-gravitational
matter fields are confined; on the other hand, gravity enters the
extra spatial dimensions allowed to be much larger than
$l_\mathrm{P}\sim 10^{-33}~\mathrm{cm}$ \cite{Ark-Dim-Dva:1998:}.
Gravity can be localized near the brane even with an infinite size
extra dimension with the warped spacetime satisfying the 5D
Einstein equations with a negative cosmological constant, as shown
by Randall and Sundrum \cite{Ran-Sun:1999:}. Consequently, an
arbitrary energy-momentum tensor could be allowed on the brane
with effective 4D Einstein equations \cite{Shi-Mae-Sas:2000:}.

The Randall--Sundrum model gives standard 4D Einstein equations in
the low-energy limit, but significant deviations occur at very
high energies, e.g., in the vicinity of compact objects, such as
black holes \cite{Dad-etal:2000:} or neutron stars
\cite{Ger-Maa:2001:}. There are high-energy effects of local
origin influencing pressure of matter and non-local effects of
``back-reaction'' origin arising from the influence of the Weyl
curvature of the bulk space on the brane.  The combination of the
high-energy (local) and bulk stress (non-local) effects alters
significantly the matching problems on the brane in comparison
with the standard 4D gravity \cite{Maa:2004:}, and the stresses
induced by the bulk gravity imply that the matching conditions do
not have unique solution on the brane \cite{Maa:2004:}. The 5D
Weyl tensor is needed as a minimum condition for uniqueness, but
no exact 5D solution is known recently \cite{Ger-Maa:2001:}.

Assuming a spherically symmetric metric induced on the brane,
solutions of the effective 4D Einstein equations could be found
relatively easily. The vacuum (in both the bulk and brane)
solutions represent black holes (naked singularities) of the
Reissner--Nordstr\"{o}m (RN) type endowed with a brany tidal
charge parameter $b$, describing the bulk tidal stress effect on
the black hole structure \cite{Dad-etal:2000:}. In the brany RN
metric, the tidal charge $b$ stands instead of the charge
parameter $Q^2$ of the standard RN
spacetimes~\cite{Mis-Tho-Whe:1973:Gra:}; $b$ can be both positive
and negative, but there are indications that $b<0$ should properly
represent the effects of the bulk space Weyl tensor
\cite{Dad-etal:2000:,Sas-Shi-Mae:2000:,Maa:2004:}.\footnote{The
stationary and axially symmetric vacuum solutions describing
rotating black holes (naked singularities) are found to be of the
Kerr--Newman type endowed again with the tidal charge reflecting
the effect of the bulk stresses \cite{Ali-Gum:2005:}.}

In the case of spherically symmetric neutron stars (or quark
stars) simplified by the uniform energy density profile, the local
high-energy and non-local bulk graviton stress effects make the
matching conditions unambitious, since we do not know the 5D Weyl
tensor. Note that the uniform energy density distribution model of
neutron stars could, in principle, be used in order to roughly
determine the neutron star spacetime structure, since in realistic
models with a variety of realistic equations of state, the energy
density profile happens to be nearly uniform in most of the
neutron star interior
\cite{Web-Gle:1992:ASTRJ2:upr:,Gle:1997:CompactStars:}.

Two different exact exterior solutions related to the uniform
density interior are known, both having asymptotically
Schwarzschild character \cite{Ger-Maa:2001:}. The first one is of
the known RN type with the tidal charge and mass parameters being
influenced by the energy density of the star and the brane
tension. The second one is more complicated, with metric
components being explicitly dependent on the energy density and
brane tension \cite{Ger-Maa:2001:}. Quite recently, the solution
of the RN type was used in the weak-field limit in order to obtain
restrictions on the brany parameter and the brane tension from the
standard tests on the Mercury perihelion precession, deflection of
light near the edge of Sun and the radar echo experiments
\cite{Boh-etal:2008:SolarSys:}. Here, we also restrict our
investigations to the metric of the RN type in order to obtain an
idea of the tidal charge relevance in a strong gravitational field
near neutron stars (or black holes) and to find restrictions on
the allowed values of the brany parameter. The choice of the RN
type of the uniform star brany solutions is supported by a recent
study of static exteriors of non-static brany stars showing that
spherical non-rotating star solutions admit only Schwarzschild or
RN type exterior \cite{PdeLeon:2008:}. Since the junction
conditions of the internal and external neutron star spacetimes
relate the tidal charge to the brane tension and the energy
density (mass and radius) of the neutron star, we can, in
principle, also put restrictions on the allowed values of the
brane tension \cite{Ger-Maa:2001:}.

There is a variety of papers related to both weak-field
\cite{Kee-Pet:2006:brane-lensing:} and strong-field
\cite{Boz:2005:,Eir:2005:,Whi:2005:} optical lensing phenomena in
the braneworld black hole spacetimes. The optical phenomena
connected to the accretion disc radiation could also reflect in a
relevant way the influence of the tidal charge
\cite{Sch-Stu:2007:RAGtime8and9CrossRef:SpBranKBH,
Sch-Stu:2007:RAGtime8and9CrossRef:OEBraK}. Here we focus our
attention on the strong-field phenomena closely related to the
perihelion precession effect discussed in the weak-field limit
\cite{Boh-etal:2008:SolarSys:}.

High-frequency quasiperiodic oscillations (QPOs) observed in
Galactic low-mass binary systems, namely, in both the black hole
\cite{Rem-McCli:2006:ARASTRA:,Kli:2004:,Kli:2000:ARASTRA:,Ter-Abr-Klu:2005:ASTRA:QPOresmodel}
and neutron star binary systems
\cite{Bel-Men-Hom:2005:ASTRA:,Bel-Men-Hom:2007:astro-ph/0702157:,Bar-Oli-Mil:2005:MONNR:,Tor:2007:ASTRA:RevTwPk6},
or in extragalactic black hole sources \cite{Str:2007:astro-ph/0701390:,Lac-Cze-Abr:2006:astro-ph0607594:,Asc-etal:2004:ASTRA:,Ter:2005:astro-ph0412500:}, 
represent such an interesting phenomenon in the strong field
regime enabling estimates of the tidal charge on the basis of
orbital models. It is widely accepted that the high-frequency QPOs
are related to the orbital motion in the inner part of an
accretion disc around a compact object. There are strong
indications that some resonant phenomena could be relevant in both
black hole systems where twin peak QPOs with fixed frequencies and
their ratios are observed
\cite{Ter-Abr-Klu:2005:ASTRA:QPOresmodel}, and neutron star
systems with wide spread of frequencies and their ratios where the
twin peak QPOs are cumulated around small number ratios
\cite{Tor-etal:2008:AA:,Tor-etal:2008:AA2:}. The hypothetical
resonances of the orbital epicyclic oscillations representing
plausible models of QPOs were recently studied for brany Kerr
black holes \cite{SK:2008:PhysRevD:}. Here, we study these models
as related to the brany neutron stars described approximately by
the RN spacetimes, i.e., neglecting the effects related to
rotation of neutron stars. We focus our attention to the
relativistic precession model \cite{Ste-Vie:1998:ASTRJ2L:} that
fits qualitatively well in the wide spread data given by
observations of a variety of neutron star binary systems
\cite{Bel-Men-Hom:2005:ASTRA:,Bel-Men-Hom:2007:astro-ph/0702157:,Bar-Oli-Mil:2005:MONNR:,Tor:2007:ASTRA:RevTwPk6}.  
However, recently it has been discussed in the case of the atoll
source 4U~1636$-$53
\cite{Tor-etal:2008:AA:,Tor-etal:2008:AA2:,Tor-Bak-Stu-Cec:2007::prep}
and some other atoll sources
\cite{Tor-etal:2007:RAGtime8and9CrossRef:MutRelkHzQPO} that the
resonant phenomena should be relevant even in the framework of the
relativistic precession (RP) model. We, thus, present a first
study of strong-field phenomena for brany neutron stars that could
be compared  with the weak-field results (perihelion precession,
radar echo delay and light deflection) based on the RN solutions
used to describe approximately the spacetime in the vicinity of
Sun \cite{Boh-etal:2008:SolarSys:}. In order to obtain rough
restrictions on both the brany tidal charge and brane tension, we
use an ensemble of typical neutron star atoll sources and the
special cases of the Z-sources Sco X-1 and Circinus X-1 that
demonstrate a wide variety of QPOs.

In~\sref{sec:dve}, the effective 4D Einstein equations are
presented, and the braneworld neutron star spacetime of the RN
type (approximated by the uniform energy density distribution
\cite{Ger-Maa:2001:}) is briefly described. In~\sref{sec:tri}, the
circular geodesics are discussed. In~\sref{sec:ctyri}, frequencies
of the orbital epicyclic oscillations are given; their radial
profiles are discussed, and the resonance conditions are briefly
described for some frequency ratio functions modeling twin peak
QPOs. In~\sref{sec:pet}, the RP model is used to fit data observed
in the representative ensemble of neutron star X-ray binaries
exhibiting QPOs, and the restrictions on the tidal charge and
brane tension are given. In~\sref{sec:sest}, concluding remarks
are presented.

\section{\label{sec:dve}Gravitational field equations on the brane and the exterior spacetime to the uniform density configurations}

In the 5D warped space models involving a non-compact extra
dimension, the gravitational field equations in the bulk can be
expressed in the form \cite{Shi-Mae-Sas:2000:,Dad-etal:2000:}
\begin{equation}
 \widetilde{G}_{AB}=\widetilde{k}^2[-\widetilde\Lambda g_{AB}+\delta(\chi)
 (-\lambda g_{AB}+T_{AB})],\label{beq1}
\end{equation}
where the fundamental 5D Planck mass $\widetilde{M}_{\mathrm{P}}$
enters via $\widetilde{k}^2=8\pi/\widetilde{M}_{\mathrm{P}}^3$,
$\lambda$ is the brane tension, and $\widetilde\Lambda$ is the
bulk cosmological constant.

The effective gravitational field equations induced on the brane
are determined by the bulk field equation \eref{beq1}, the
Gauss--Codazzi equations and the generalized matching Israel
conditions with $Z_2$ symmetry. They can be expressed as modified
standard Einstein's equations containing additional terms
reflecting bulk effects onto the brane \cite{Shi-Mae-Sas:2000:}:
\begin{equation}
 G_{\mu\nu}=-\Lambda g_{\mu\nu}+k^2 T_{\mu\nu} + \widetilde{k}^2
 S_{\mu\nu} -E_{\mu\nu},
\end{equation}
where $k^2=8\pi/M_{\mathrm{P}}^2$, with $M_{\mathrm{P}}$ being the
brany Planck mass. The relations of the energy scales and
cosmological constants are given in the form
\begin{equation}
 M_{\mathrm{P}}=\sqrt{\frac{3}{4\pi}}\left(\frac{\widetilde{M}_{\mathrm{P}}^2}{\sqrt{\lambda}}\right)\widetilde{M}_{\mathrm{P}};
 \quad \Lambda=\frac{4\pi}{\widetilde{M}_{\mathrm{P}}^3}\left[\widetilde\Lambda+\left(\frac{4\pi}
 {3\widetilde{M}_{\mathrm{P}}^3}\right)\lambda^2\right].
\end{equation}
Requirement of zero cosmological constant on the brane
($\Lambda=0$) implies the relation
\begin{equation}
 \widetilde\Lambda=-\frac{4\pi\lambda^2}{3\widetilde{M}_{\mathrm{P}}^3}.
\end{equation}

Local bulk effects on the matter are determined by the ``squared
energy-momentum'' tensor $S_{\mu\nu}$ arising from the extrinsic
curvature term in the projected Einstein tensor:
\begin{equation}
 S_{\mu\nu}=\frac{1}{12}T T_{\mu\nu}-\frac{1}{4}T_\mu^{\phantom{\mu}\alpha}
 T_{\nu\alpha}+\frac{1}{24}g_{\mu\nu}\left(3T^{\alpha\beta}T_{\alpha\beta}-T^2\right),
\end{equation}
where $T_{\mu\nu}$ is the energy-momentum tensor on the brane. The
non-local bulk effects are given by the tensor $E_{\mu\nu}$
representing the bulk Weyl tensor, $\widetilde{C}_{ACBD}$,
projected onto the brane, whereas
\begin{equation}
 E_{AB}=\widetilde{C}_{ACBD}\,n^C n^D.
\end{equation}

There are two known solutions of the effective 4D Einstein
equations corresponding to the brany uniform density
configurations that could roughly represent the spacetime exterior
to neutron (quark) stars \cite{Ger-Maa:2001:}. Here, we focus our
attention on the solution of the Reissner--Nordstr\"{o}m type that
is formally identical to the spherically symmetric black hole
solution; however, its parameters are related to the brane tension
$\lambda$ and energy density $\rho$ of the internal configuration
due to the matching conditions of the internal and external
spacetimes. The external spacetime is given in the standard
Schwarzschild coordinates by the line element:
\begin{eqnarray}
\rmd s^2&=& -A(r)\mathrm{c}^2 \rmd t^2+A^{-1}(r)\rmd r^2+r^2\rmd
\theta^2+r^2\sin^2\theta\,\rmd \phi^2,\\
A(r)&=&1-\frac{2r_{\mathrm{G}}}{r}+\frac{B}{r^2};\qquad
r_{\mathrm{G}}=\frac{\mathrm{G}M}{\mathrm{c}^2},
\end{eqnarray}
where M is the gravitational mass, and $B$ is the brany tidal
charge representing the bulk non-local effects on the 4D spacetime
structure. The matching conditions imply \cite{Ger-Maa:2001:}
\begin{equation}\label{match:cond}
B=-\frac{3\mathrm{G}M}{\mathrm{c}^2}R\left(\frac{\rho}{\lambda}\right),
\end{equation}
\begin{equation}
M=\widetilde{M}\left(1-\frac{\rho}{\lambda}\right),
\end{equation}
where $R$ is the internal configuration radius, and the effective
mass is given by
\begin{equation}
\widetilde{M}=4\pi \int\limits_0^{R}\rho^{\mathrm{eff}}(r) r^2
\rmd r,
\end{equation}
with the effective density given by
\begin{equation}
\rho^{\mathrm{eff}}=\rho
\left(1+\frac{\rho}{2\lambda}\right)+\frac{6}{\left(8\pi
\mathrm{G}\right)^2\lambda}\,\mathcal{U}.
\end{equation}
The non-local bulk gravitational effects arising from the bulk
Weyl tensor are represented by the ``dark energy'' density
$\mathcal{U}$. The standard general relativistic equations are
regained in the limit of $\lambda^{-1}\rightarrow 0$.

The internal solution puts two important restrictions on the brane
tension $\lambda$ \cite{Ger-Maa:2001:}. The first one is general
for all uniform stars and reads
\begin{equation}
\lambda \geq \left(\frac{\mathrm{G}M/\mathrm{c}^2}{R-2
\mathrm{G}M/\mathrm{c}^2}\right)\rho\,,
\end{equation}
implying $R>2\mathrm{G}M/\mathrm{c}^2$, i.e., the Schwarzschild
radius is still the relevant limit as in general relativity.

The second one represents an upper limit on compactness of the
star following from the requirement that pressure must be finite
inside the star. It takes the form \cite{Ger-Maa:2001:}
\begin{equation}\label{compact-cond}
\frac{\mathrm{G}M/\mathrm{c}^2}{R}\leq
\frac{4}{9}\left[\frac{1+5\rho/4\lambda}{\left(1+\rho/\lambda\right)^2}\right]
\end{equation}
that implies the general relativistic limit
$\left(\mathrm{G}M/\mathrm{c}^2\right)/R\leq 4/9$ for
$\lambda^{-1}\rightarrow 0$. We can see that the braneworld
high-energy corrections reduce the compactness limit of the star
and the lowest order correction reads
\begin{equation}
\frac{\mathrm{G}M/\mathrm{c}^2}{R}\leq
\frac{4}{9}\left(1-\frac{3}{4}\frac{\rho}{\lambda}\right).
\end{equation}

\section{\label{sec:tri}Circular geodesics of the external spacetime}

\begin{figure*}[p]
\begin{center}
\subfigure[][]{\label{orbity-vse:a}\includegraphics[width=.76\hsize]{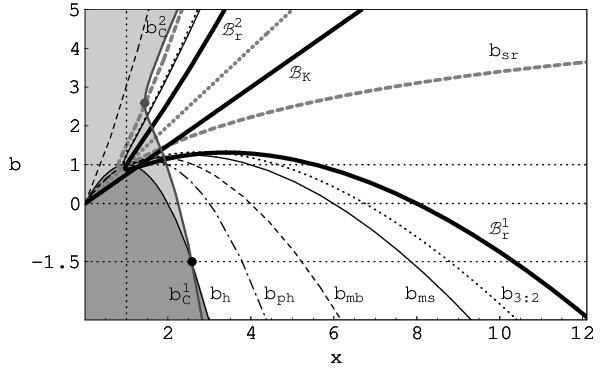}}\\
\subfigure[][]{\label{orbity-vse:b}\includegraphics[width=.76\hsize]{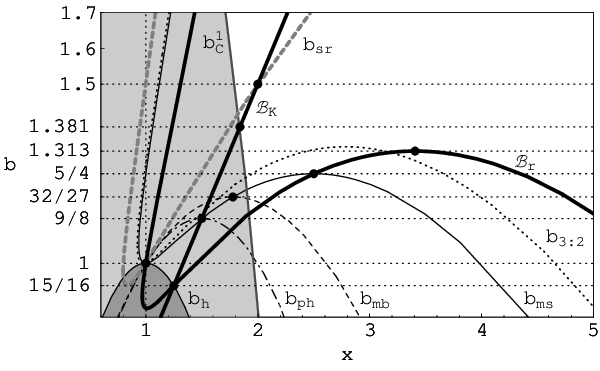}}
\end{center}
\caption{\label{orbity-vse}(a) The functions $b_{\mathrm{h}}$
(black solid line), $b_{\mathrm{ph}}$ (dashed-dotted line),
$b_{\mathrm{mb}}$ (dashed line) and $b_{\mathrm{ms}}$ (thin black
solid line) that implicitly determine the radius of the outer
event horizon, the limiting photon orbit and the marginally bound
and marginally stable circular orbits of a braneworld RN type
spacetime. The grey solid line represents the function
$b_{\mathrm{C}}$ implicitly determining the critical radius
$X_{\mathrm{C}}$ of neutron stars. The thick black solid lines
represent the functions
$\mathcal{B}_{\mathrm{K}}(x=\mathcal{X_{\mathrm{K}}})$ and
$\mathcal{B}_{\mathrm{r}}^{k}(x=\mathcal{X_{\mathrm{r}}})$,
implicitly determining the locations $\mathcal{X}_{\mathrm{K}}$
and $\mathcal{X}_{\mathrm{r}}$ of the Keplerian and radial
epicyclic frequency local extrema. Thick grey dotted line
represents the function $b=x$ determining the other limit on the
reality of circular orbits. The function $b_{\mathrm{sr}}$ that
implicitly determines the strong resonant radius
$x_{\mathrm{sr}}$, where
$\nu_{\mathrm{K}}(x,b)=\nu_{\mathrm{r}}(x,b)$
($\alpha_{\mathrm{r}}(x,b)=1$) is denoted by thick grey dashed
line. The function $b_{3:2}
$ (black dotted line) represents the radii, where the relativistic
precession resonance $\nu_{\mathrm{K}}\!:\!\left(
\nu_{\mathrm{K}}-\nu_{\mathrm{r}}\right)=3\!:\!2$ occurs. (b) The
detailed figure introduces the relevant values of the brany tidal
charge, separating the external spacetimes of different
properties.}
\end{figure*}

Considering vacuum spacetimes, the event horizons of the brany RN
metric are determined by the condition $A(r)=0$. Using geometric
units ($\mathrm{c}=\mathrm{G}=1$), the radius of the outer event
horizon is given by the relation
\begin{equation}\label{horizont}
r_+=M+\sqrt{M^2-B} \,.
\end{equation}
The horizon structure depends on the sign of the tidal charge $B$.
We see that, in contrast to its positive values, the negative
tidal charge tends to increase the horizon radius (see, e.g.,
\fref{orbity-vse}).

It is clear that the positive tidal charge acts to weaken the
gravitational field, and we have the same horizon structure as in
the usual Reissner--Nordstr\"{o}m solution. However, new
interesting features arise for the negative tidal charge. For
$B<0$, it follows from equation~\eref{horizont} that the horizon
radius \cite{Dad-etal:2000:}
\begin{equation}
r_+ > 2M ;
\end{equation}
such a situation is not allowed in the framework of general
relativity.

The event horizon does exist provided that $M^2\geq B$, where the
equality corresponds to the extreme case. For $B > M^2$ we obtain
a naked singularity spacetime. Clearly, such a situation is
impossible for negative values of the brany tidal charge $B$.
Here, we are confronted with another situation, since the vacuum
spacetime is limited to $r>R$ and at $r=R$ it has to be matched to
the internal uniform density spacetime. Clearly, the radius $R$
must satisfy the compactness condition~\eref{compact-cond}.

Note that values of the brany tidal charge $B$ are not restricted
for both positive and negative values; therefore, the external
spacetimes with $B>M^2$ (corresponding to the ``naked
singularity'' case) have to be considered too.

In the analysis of the epicyclic frequency profiles, it is useful
to relate the profiles to the photon circular geodesic and
innermost stable circular geodesic radii (or innermost bound
circular geodesic radii in the case of thick discs not considered
here), which are relevant in the discussion of properties of the
accretion discs and their oscillations. Therefore, we put the
limiting radii in an appropriate form.

Introducing the dimensionless radial coordinate
\begin{equation}\label{def-x}
x=\frac{r}{r_{\mathrm{G}}}
\end{equation}
and a dimensionless brany parameter
\begin{equation}
b=\frac{B}{r^2_{\mathrm{G}}},
\end{equation}
the metric coefficient takes the form
\begin{equation}
A(x,b)=1-\frac{2}{x}+\frac{b}{x^2}.
\end{equation}

The motion of a test particle with mass $m$ is given by the
standard geodesic equation $\mathrm{D} U^{\mu} / \rmd \tau=0$ and
the normalization condition $U_\mu U^\mu=-m^2$. The motion is
restricted to the central planes; for one particle the plane can
be chosen to be the equatorial plane
($\theta=\pi/2=\mbox{const}$). Then, the relevant radial equation
of motion takes the form
\begin{equation}
    \left(\frac{\rmd x}{\rmd \tau}\right)^2= E^2-V^2_{\mathrm{eff}}(x,L,b),
\end{equation}
where the effective potential
\begin{equation}
    V^2_{\mathrm{eff}}(x,L,b)=A(x,b)\left(m^2+ \frac{L^2}{x^2}\right).
\end{equation}
There are two constants of motion related to the spacetime
symmetries: $E/m$ is the specific energy; $L/m$ is the specific
angular momentum. The circular geodesic orbits (determined by
$\partial V_{\mathrm{eff}} / \partial x =0$) have the specific
energy and the specific angular momentum given by
\begin{eqnarray}
    \frac{E}{m}&=&\frac{x^2-2x+b}{x\left(x^2-3x+2b\right)^{1/2}}\,,\label{energie-bezrozm}\\
    \frac{L}{m}&=&\pm\frac{x\sqrt{x-b}}{\left(x^2-3x+2b\right)^{1/2}}\,.\label{moment-bezrozm}
\end{eqnarray}

From equations~\eref{energie-bezrozm} and \eref{moment-bezrozm} we
can see that the circular orbits can exist from infinity down to
the radius of the limiting circular photon orbit, determined by
the condition
\begin{equation}\label{rce-foton}
    x^2-3 x +2b=0 ,
\end{equation}
where $E/m\rightarrow \infty$ and $L/m\rightarrow \pm\infty$, but
the impact parameter $I=L/E$ remains finite. For ``naked
singularity'' RN spacetimes, the restriction
\begin{equation}
    x\geq b
\end{equation}
can be relevant.

The radius of the marginally bound orbits with $E^2=m^2$ is given
by the largest root of the equation
\begin{equation}\label{rce-mezni-vaz}
    x\left(4 x -x^2-4b\right)+b^2=0,
\end{equation}
while the stable circular orbits are restricted by the condition
\begin{equation}\label{rce-mezni-stab}
 x\left(6 x - x^2-9b\right)+4b^2 \leq 0 .
\end{equation}
The outer event horizon $x_{\mathrm{h}}(b)$ is implicitly
determined by the relation
\begin{equation}\label{hor-impl}
b=b_{\mathrm{h}}(x)\equiv x(2-x),
\end{equation}
the radius of the photon circular orbit $x_{\mathrm{ph}}(b)$ is
given by
\begin{equation}\label{fot-impl}
b=b_{\mathrm{ph}}(x)\equiv \frac{x}{2}(3-x),
\end{equation}
the radius of the marginally bound orbit $x_{\mathrm{mb}}(b)$ is
given by
\begin{equation}\label{mez-vazana-impl}
b=b_{\mathrm{mb}}(x)\equiv x\left(2\mp \sqrt{x}\right),
\end{equation}
and the radius of the marginally stable orbit $x_{\mathrm{ms}}(b)$
is given by
\begin{equation}\label{ms-impl}
b=b_{\mathrm{ms}}(x)\equiv \frac{x}{8}\left(9\mp
\sqrt{16x-15}\right).
\end{equation}
The upper sign in \eref{mez-vazana-impl} and \eref{ms-impl} is
relevant for the black hole spacetimes in the region above the
outer horizon, while both signs are relevant for the naked
singularity spacetimes.

Now, we have to put the limits on the applicability of the vacuum
spacetime using the limit condition on the compactness of the
uniform internal configuration given by the
relation~\eref{compact-cond}. We shall give the limits in terms of
the dimensionless brany tidal charge $b$ that can be expressed in
the form
\begin{equation}
b=-3\,\frac{R}{r_{\mathrm{G}}}\left(\frac{\rho}{\lambda}\right)
=-3X\left(\frac{\rho}{\lambda}\right).
\end{equation}
Clearly, negative (positive) values of $b$ correspond to positive
(negative) values of the brany tension $\lambda$.

The compactness relation~\eref{compact-cond} then reads
\begin{equation}\label{compact-cond-dosazeni}
4
-\left(9+\frac{5}{3}b\right)\left(\frac{r_{\mathrm{G}}}{R}\right)+6b
\left(\frac{r_{\mathrm{G}}}{R}\right)^2-b^2\left(\frac{r_{\mathrm{G}}}{R}\right)^3\geq
0\,,
\end{equation}
and in the linear form
\begin{equation}
\left(\frac{r_{\mathrm{G}}}{R}\right)\leq \frac{4}{9-b}\,.
\end{equation}
The equality in relation~\eref{compact-cond-dosazeni} determines
the critical radius $R_{\mathrm{C}}$ representing the limit on the
radius of neutron stars. We have to restrict our studies to the
region $r>R_{\mathrm{C}}$. The radii
$X_{\mathrm{C}}=R_{\mathrm{C}}/r_{\mathrm{G}}$ are illustrated in
\fref{orbity-vse}, and are given (in an implicit form) by the
relation
\begin{equation}\label{krit:pol}
   b = b_{\mathrm{C}}(x) \equiv \frac{1}{6}\left(18x-5x^2\mp x^{3/2} \sqrt{25
    x-36}\right).
\end{equation}

The functions $b_{\mathrm{h}}(x)$, $b_{\mathrm{ph}}(x)$,
$b_{\mathrm{mb}}(x)$, $b_{\mathrm{ms}}(x)$ and $b_{\mathrm{C}}(x)$
are illustrated in \fref{orbity-vse}. It is evident that the
positive tidal charge will play the same role in its effect on the
circular orbits as the electric charge in the
Reissner--Nordstr\"{o}m spacetime -- the radius of the circular
photon orbit, as well as the radii of the innermost bound and the
innermost stable circular orbits shift inwards as the positive
tidal charge increases. For the negative tidal charge the limiting
photon orbit, the innermost bound and the innermost stable
circular orbits shift outwards as the absolute value of $b$
increases \cite{Dad-etal:2000:,Ali-Gum:2005:}. Of special interest
is the behaviour of $b_{\mathrm{C}}(x)$ putting limits on validity
of the internal uniform density configuration. For $b=0$, the
limit is $X_{\mathrm{C}}=9/4$, and it falls with $b>0$ growing up
to $b\doteq 2.6$, where it starts to grow again. On the other
hand, for $b<0$ the critical radius $X_{\mathrm{C}}$ slowly grows
with $b$ descending to higher negative values, but for $b=-1.5$,
there is $X_{\mathrm{C}}=x_{\mathrm{h}}$, and the function
$b_{\mathrm{C}}(x)$ loses its relevance leaving $X>x_{\mathrm{h}}$
as the only basic restriction on the surface radius of the neutron
star.

The function $b_{\mathrm{h}}(x)$ has a local maximum at $x=1$
where $b=1$, corresponding to the extreme case. This local maximum
coincides with the local minimum of the function
$b_{\mathrm{ms}}(x)$ at $x=1$. The function $b_{\mathrm{ph}}(x)$
has a local maximum at $x=3/2$ for $b=9/8$ corresponding to the
naked singularity spacetimes. The local maximum of the function
$b_{\mathrm{mb}}(x)$ is located at $x=16/9$ for $b=32/27$; the
function $b_{\mathrm{ms}}(x)$ has a local maximum situated at
$x=5/2$ for $b=5/4$ (this local maxima appear again only for naked
singularities); the function $b_{\mathrm{C}}(x)$ crosses
$b_{\mathrm{h}}(x)$ at $b=-1.5$ (see \fref{orbity-vse}). Clearly,
due to the compactness limit $b_{\mathrm{C}}(x)$, the photon
circular orbits are relevant in the spacetimes with $b \leq 1.017$
only. Further, we immediately see that for $b>0$, the marginally
stable orbits are well defined in the black hole type spacetimes
($b \leq 1$) and in naked singularity type spacetimes with $b \leq
5/4$ and $b > 9/4$. In the spacetimes with $5/4<b\leq 9/4$, the
stable orbits can be located at all $x\geq X_{\mathrm{C}}$, since
$x_{\mathrm{ms}}<X_{\mathrm{C}}$ (and
$x_{\mathrm{ms}}=X_{\mathrm{C}}$ for $b=9/4$) in such vacuum RN
spacetimes.

\section{\label{sec:ctyri}Epicyclic oscillations of Keplerian motion}

It is well known that for oscillations of both thin Keplerian
\cite{Kat-Fuk-Min:1998:BHAccDis:,Klu-Abr:2000:ASTROPH:} and
toroidal discs \cite{Rez-etal:2003:MONNR:} around neutron stars
(black holes) the orbital Keplerian frequency $\nu_{\mathrm{K}}$
and the related radial and vertical epicyclic frequencies
$\nu_{\mathrm{r}}$ and $\nu_{\theta}$ of geodetical quasicircular
motion are relevant and observable directly or through some
combinational frequencies
\cite{Ter-Abr-Klu:2005:ASTRA:QPOresmodel,
Tor-Stu:2005:RAGtime6and7:CrossRef,Ter-Stu:2005:ASTRA:,
Stu-Kot-Tor:2007:RAGtime8and9CrossRef:MrmQPO}. Of course, for
extended tori, the eigenfrequencies of their oscillations are
shifted from the epicyclic frequencies in dependence on the
thickness of the torus
\cite{Sra:2005:ASTRN:,Bla-etal:2006:ASTRJ2:}. Similarly, due to
non-linear resonant phenomena, the oscillatory eigenfrequencies
could be shifted from the values corresponding to the resonant
geodetical epicyclic frequencies in dependence on the oscillatory
amplitude \cite{Lan-Lif:1976:Mech:,Nay-Moo:1979:NonOscilations:}.
It is expected that shift of this kind is observed in neutron star
systems
\cite{Abr-etal:2005:RAGtime6and7:CrossRef,Abr-etal:2005:ASTRN:},
while in microquasars, i.e., binary black hole systems, the
observed frequency scatter is negligible and the geodetical
epicyclic frequencies should be relevant.

In the case of the external braneworld neutron star spacetime of
the RN type with the brany tidal charge $b$, the formulae of the
test particle geodetical circular motion and its epicyclic
oscillations, obtained in \cite{Ali-Gal:1981:}, could be directly
applied (putting $a=0$ and $Q^2=b$). We can write the following
relations for the orbital and epicyclic frequencies:
\begin{eqnarray}
\nu_{\mathrm{K}}&=&\frac{1}{2\pi}\left(\frac{\mathrm{G }M
}{r_{\mathrm{G}}^3}\right)^{1/2}\frac{1}{x^{3/2}}\left(1-\frac{b}{x}\right)^{1/2}=
\frac{1}{2\pi}\frac{\mathrm{c}^3}{\mathrm{G}M}\frac{1}{x^{3/2}}\left(1-\frac{b}{x}\right)^{1/2},\label{def-Kep}\\
\nu^2_{\theta}&=&\alpha_{\theta}\nu_{\mathrm{K}}^2,\\
\nu^2_{\mathrm{r}}&=&\alpha_{\mathrm{r}}\nu_{\mathrm{K}}^2,\label{def-epi-rad}
\end{eqnarray}
where the dimensionless quantities determining the epicyclic
frequencies are given by
\begin{eqnarray}
\alpha_{\theta}&=&1,\\
\alpha_{\mathrm{r}}&=&\left(1-\frac{b}{x}\right)^{-1}\left(1-\frac{6}{x}+\frac{9b}{x^2}-\frac{4b^2}{x^3}\right)\label{alfa:r},
\end{eqnarray}
so that $\nu_{\mathrm{K}}(x,b)=\nu_{\theta}(x,b)$ due to the
spherical symmetry of the spacetime.

In the field of brany RN black holes (neutron stars) with $b \leq
1$, there is (see \fref{frkv-obecne})
\begin{equation}
\nu_{\mathrm{K}}(x,b)>\nu_{\mathrm{r}}(x,b).
\end{equation}
However, this statement is not generally correct in the case of
brany RN naked singularities (neutron stars) with $b>1$. For $b
> 1.42$, the case $\nu_{\mathrm{K}}(x,b)<\nu_{\mathrm{r}}(x,b)$ is even
possible in some region of the external (vacuum) spacetime.

Now we can discuss the behaviour of the fundamental orbital
frequencies for Keplerian motion in the field of both brany RN
black hole and brany RN naked singularity spacetimes, i.e., we
consider external neutron star spacetimes with tidal charge $b
\leq 1$ and $b>1$. (Note that $b<0$ always corresponds to the
black hole spacetime behaviour.)

We express the frequency as
$\nu\,[\mathrm{Hz}]\,10\,\mathrm{M}_{\odot}/M$ in every
quantitative plot of frequency dependence on radial coordinate
$x$, i.e., displayed value is the frequency relevant for a central
object with a mass of $10\,\mathrm{M}_{\odot}$, which could be
simply rescaled for another mass just by dividing the displayed
value by the respective mass in units of ten solar masses.

\subsection{Properties of the Keplerian and
radial epicyclic frequency}

First, it is important to find the range of relevance for the
functions $\nu_{\mathrm{K}}(x,b)$ and $\nu_{\mathrm{r}}(x,b)$
above the critical radius $R_{\mathrm{C}}$ of neutron stars. For
completeness we consider their behaviour above the event horizon
$x_{\mathrm{h}}$ for black holes (neutron stars), and above the
ring singularity located at $x = 0$ for naked singularities;
recall that in the external spacetimes with $b<-1.5$, the
restriction by $x>X_{\mathrm{C}}(b)$ is irrelevant, but the
horizon condition $x>x_{\mathrm{h}}(b)$ keeps its relevance.

The circular geodesics can exist for $x>x_{\mathrm{ph}}(b)$, where
$x_{\mathrm{ph}}(b)$ is implicitly determined by the equation
\eref{rce-foton}. Stable circular geodesics, relevant for the
Keplerian, thin accretion discs, exist for $x>x_{\mathrm{ms}}(b)$,
where $x_{\mathrm{ms}}(b)$ is determined (in an implicit form) by
the relation \eref{rce-mezni-stab}, which coincides with the
condition
\begin{equation}\label{podm-mez}
    \alpha_{\mathrm{r}}(x,b)=0 .
\end{equation}
For toroidal, thick accretion discs the unstable circular
geodesics can be relevant in the range $x_{\mathrm{mb}} \leq
x_{\mathrm{in}} < x < x_{\mathrm{ms}}$, being stabilized by
pressure gradients in the tori
\cite{Jar-Abr-Pac:1980:ACTAS:,Pac-Wii:1980:ASTRA:}. The radius of
the marginally bound circular geodesic $x_{\mathrm{mb}}$,
implicitly determined by the equation \eref{rce-mezni-vaz}, is the
lower limit for the inner edge of thick discs
\cite{Koz-Jar-Abr:1978:ASTRA:,Kro-Haw:2002:ASTRJ2:}.

\begin{figure*}[t]
\begin{center}
\includegraphics[width=.8\hsize]{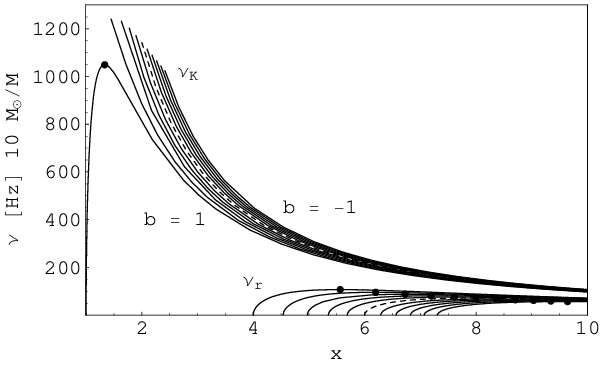}
\end{center}
\caption{\label{frkv-obecne}The behaviour of the Keplerian
frequency $\nu_{\mathrm{K}}$ and radial epicyclic frequency
$\nu_{\mathrm{r}}$ in the field of braneworld RN black hole
(neutron star) spacetimes with various values of the tidal charge
parameter $b$. The curves are spaced by $0.2$ in $b$, and they are
plotted from the outer event black hole horizon $x_{\mathrm{h}}$.
The dashed lines represent Schwarzschild spacetime with zero tidal
charge.}
\end{figure*}

Clearly, the Keplerian orbital frequency is well defined up to $x
= x_{\mathrm{ph}}$. However, $\nu_{\mathrm{r}}$ is well defined,
if $\alpha_{\mathrm{r}}\geq 0$, i.e., at $x\geq x_{\mathrm{ms}}$,
and $\nu_{\mathrm{r}}=0$ at $x_{\mathrm{ms}}$. From
\fref{frkv-obecne}, we can conclude that not only the radial
epicyclic frequency but even the Keplerian frequency can have a
maximum located above the outer event black hole horizon. In the
following subsection, we will discuss whether the maximum of
$\nu_{\mathrm{K}}(x,b)$ could be located above the marginally
stable or the limiting photon circular orbit.

\subsubsection{Local extrema of the Keplerian frequency.}

We denote by $\mathcal{X}_{\mathrm{K}}$ the local extrema of the
Keplerian frequency $\nu_{\mathrm{K}}(x,b)$ given by the condition
\begin{equation}\label{podm-max-Kep}
    \frac{\partial \nu_{\mathrm{K}}}{\partial x}=0.
\end{equation}
From \eref{def-Kep}, we find that after introducing $'$ as $\rmd
/\rmd x$ the corresponding derivative is given by
\begin{equation}\label{der-Kepl}
\nu_{\mathrm{K}}'=\frac{1}{2\pi}\sqrt{\frac{\mathrm{G}
M}{r_{\mathrm{G}}^3}}\frac{4b-3x}{2x^3\sqrt{x-b}}=
\frac{4b-3x}{2x(x-b)}\nu_{\mathrm{K}},
\end{equation}
and relation \eref{podm-max-Kep} implies that the Keplerian
frequency has a local extremum located at
\begin{equation}\label{}
\mathcal{X}_{\mathrm{K}}= \frac{4}{3}b .
\end{equation}
The second derivative at $x=\mathcal{X}_{\mathrm{K}}$,
\begin{equation}
\nu_{\mathrm{K}}''=-\frac{1}{2\pi}\sqrt{\frac{\mathrm{G}
M}{r_{\mathrm{G}}^3}}\frac{81\sqrt{3}}{128\, b^3 \sqrt{b}}\,,
\end{equation}
is always negative; thus, the Keplerian frequency has a local
maximum at $x=\mathcal{X}_{\mathrm{K}}$.

For $b=15/16$, the maximum of the Keplerian frequency radial
profile is situated exactly at the radius coinciding with the
radius of the outer event horizon
$\mathcal{X}_{\mathrm{K}}=x_{\mathrm{h}}=5/4$ (see figures
\ref{orbity-vse:b} and \ref{ex-kep-rezy}). For $b=9/8$, the
maximum of the Keplerian frequency is located at the circular
photon orbit $\mathcal{X}_{\mathrm{K}}=x_{\mathrm{ph}}=3/2$ (this
case corresponds to the naked singularity spacetime).

We can conclude that for brany parameter from the interval
\begin{equation}
15/16<b<9/8
\end{equation}
the Keplerian frequency radial profile has its maximum located
between the outer event horizon and the circular photon orbit:
\begin{equation}
x_\mathrm{h}<\mathcal{X}_{\mathrm{K}}< x_{\mathrm{ph}},
\end{equation}
i.e., it is physically irrelevant there.

Generally, the maximum of the Keplerian frequency is physically
irrelevant for all the brany RN neutron stars (of the black hole
type with $b\leq 1$), since it could never be located above the
circular photon orbit $x_{\mathrm{ph}}$ (and the marginally stable
orbit $x_{\mathrm{ms}}$). For spacetimes with $b\leq 1$, the
Keplerian frequency is a monotonically decreasing function of
radius in astrophysically relevant radii above the photon orbit,
$x>x_{\mathrm{ph}}$. The situation is clearly illustrated in
\fref{orbity-vse:b}. For the RN naked singularity type spacetimes
with $b < 1.381$ the extrema of $\nu_{\mathrm{K}}$ must be hidden
under the neutron star surface. The local extrema of
$\nu_{\mathrm{K}}$ can appear in the external spacetime to the
internal uniform density configuration for $b>1.381$ (see also
\fref{ex-kep-rezy}).

\subsubsection{Local extrema of the radial epicyclic frequency.}

The local extrema of the radial epicyclic frequency
$\mathcal{X}_{\mathrm{r}}$ are given by the condition
\begin{equation}\label{podm-max-epi}
    \frac{\partial \nu_{\mathrm{r}}}{\partial x}=0.
\end{equation}
Using~\eref{def-epi-rad}, the derivative reads
\begin{eqnarray}\label{der-epi}
\nu_{\mathrm{r}}'&=&\sqrt{\alpha_{\mathrm{r}}}\left(\nu_{\mathrm{K}}'+\frac{\alpha_{\mathrm{r}}'}{2
\alpha_{\mathrm{r}}}\nu_{\mathrm{K}}\right),\\
\alpha_{\mathrm{r}}'&=&-\frac{8b}{x^3}+\frac{5}{x^2}+\frac{1-x}{(x-b)^2}+\frac{1}{x-b},
\end{eqnarray}
where $\nu_{\mathrm{K}}'$ is given by~\eref{der-Kepl}.
Relations~\eref{podm-max-epi} and~\eref{der-epi} imply the
condition determining extrema $\mathcal{X}_{\mathrm{r}}(b)$ of the
radial epicyclic frequency profile:
\begin{equation}\label{podm-max-epi-impl}
\alpha_{\mathrm{r}}'=-\frac{2\nu_{\mathrm{K}}'}{\nu_{\mathrm{K}}}\,
\alpha_{\mathrm{r}}(x,b).
\end{equation}
In \fref{orbity-vse}, we show curves $\mathcal {B}_\mathrm{r}^k
(x=\mathcal{X}_{\mathrm{r}})$, $k\in\{1,2\}$, implicitly
determined by relation~\eref{podm-max-epi-impl}; index $k$ denotes
different branches of the solution of~\eref{podm-max-epi-impl}:
\begin{equation}
\mathcal{B}_{\mathrm{r}}^k=\frac{x}{16}\left(15\mp
\sqrt{32x-31}\right).
\end{equation}
For all possible values of the brany parameter $b\leq 1$, the
radial epicyclic frequency $\nu_{\mathrm{r}}$ has one local
maximum for braneworld RN neutron stars (black holes) that is
always located above the marginally stable orbit $x_{\mathrm{ms}}$
(see \fref{orbity-vse}).

\begin{figure*}[p]
\begin{center}
\subfigure[][]{\includegraphics[width=.31\hsize]{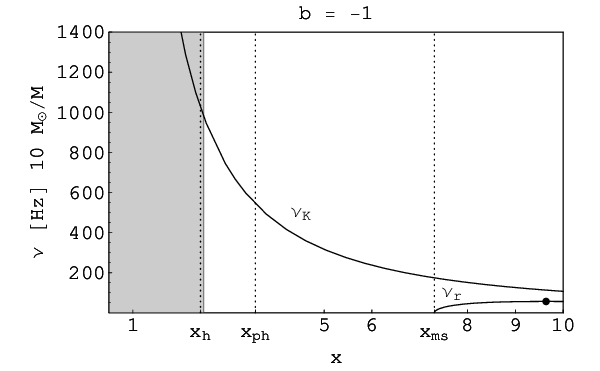}}\quad
\subfigure[][]{\includegraphics[width=.31\hsize]{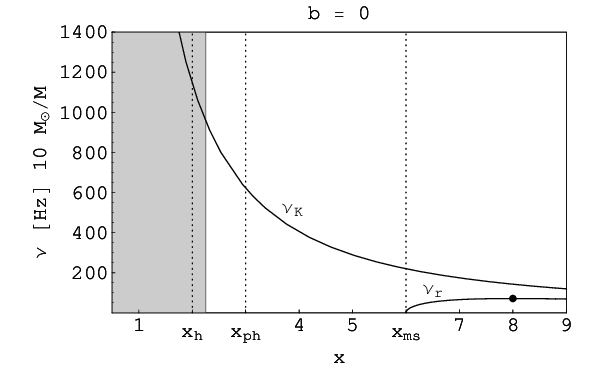}}\quad
\subfigure[][]{\includegraphics[width=.31\hsize]{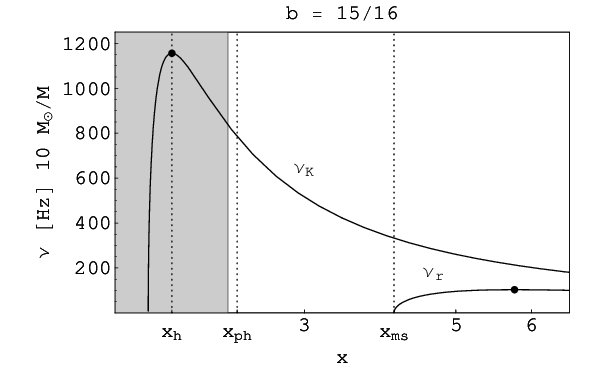}}\\
\subfigure[][]{\includegraphics[width=.31\hsize]{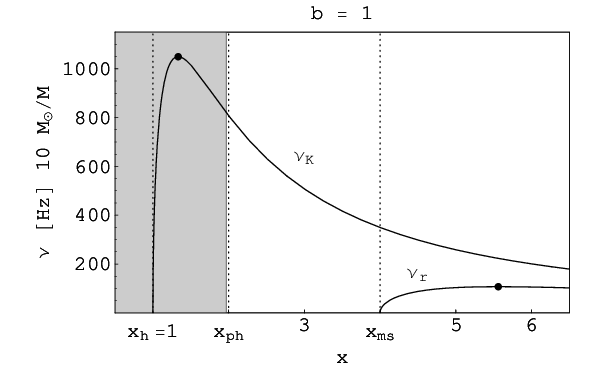}}\quad
\subfigure[][]{\includegraphics[width=.31\hsize]{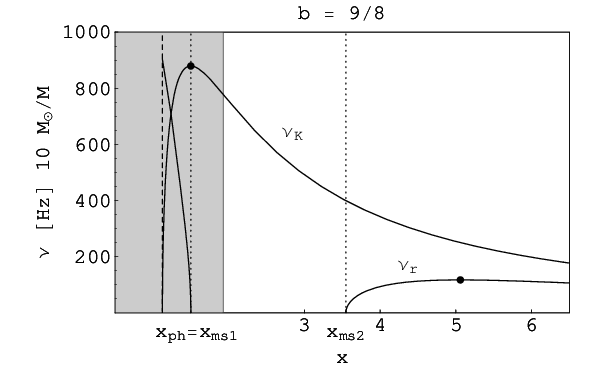}}\quad
\subfigure[][]{\includegraphics[width=.31\hsize]{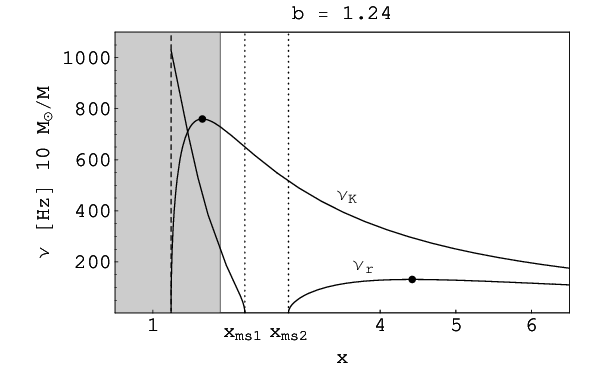}}\\
\subfigure[][]{\includegraphics[width=.31\hsize]{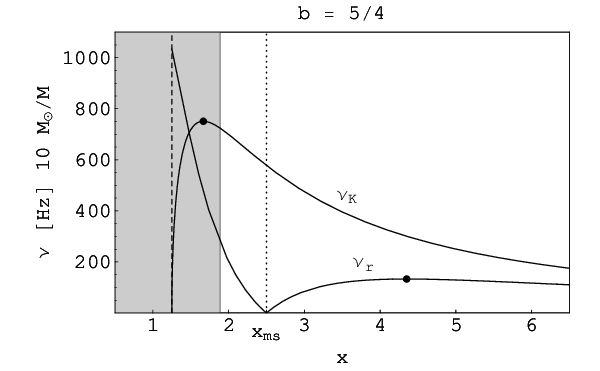}}\quad
\subfigure[][]{\includegraphics[width=.31\hsize]{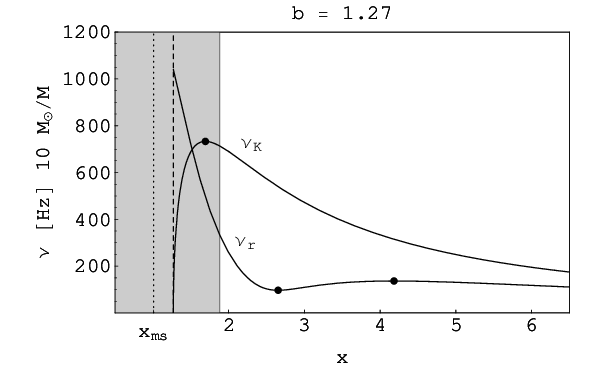}}\quad
\subfigure[][]{\includegraphics[width=.31\hsize]{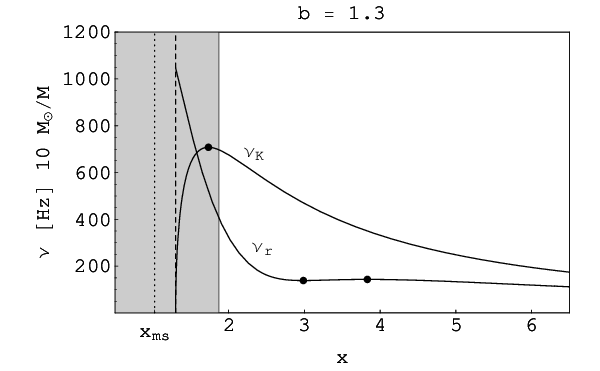}}\\
\subfigure[][]{\includegraphics[width=.31\hsize]{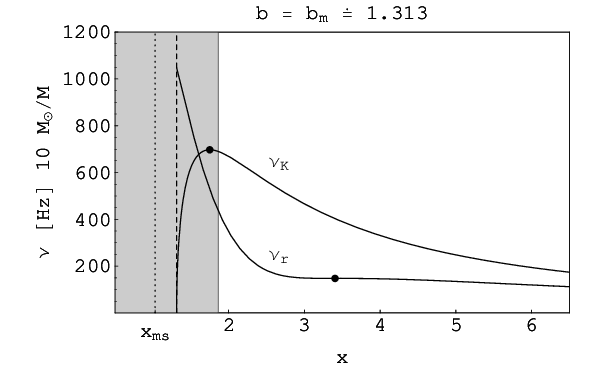}}\quad
\subfigure[][]{\includegraphics[width=.31\hsize]{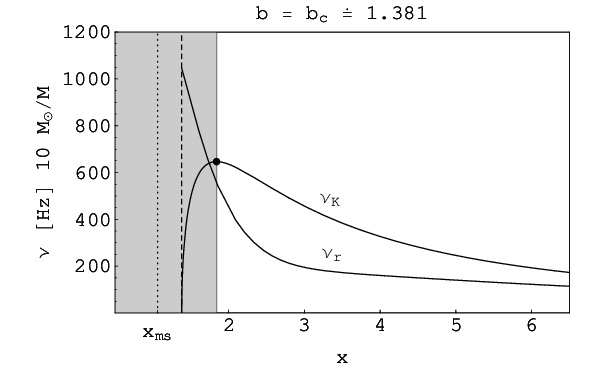}}\quad
\subfigure[][]{\includegraphics[width=.31\hsize]{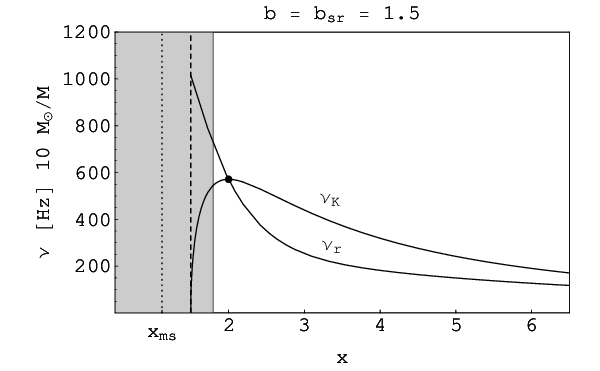}}
\end{center}\caption{\label{ex-kep-rezy}The
Keplerian and radial epicyclic frequency profiles for some
representative values of the brany parameter $b$ in braneworld RN
black hole (neutron star) (a)\,--\,(c), the extreme braneworld RN
black hole case (d), RN naked singularity spacetimes (e)\,--\,(l).
The radius of the outer event horizon $x_\mathrm{h}$, the limiting
photon orbit $x_\mathrm{ph}$ and the marginally stable
$x_\mathrm{ms}$ circular orbits are denoted by a dotted vertical
line (if these orbits exist at the given spacetime). Note that the
condition $x>b$ has to be satisfied, the dashed line in
(e)\,--\,(l) represents the radius where $x=b$. The grey solid
line denotes the critical radius $X_{\mathrm{C}}$ representing the
limit on the radius of neutron stars. So the grey area is
astrophysically irrelevant. (l) is the special case with
$\nu_{\mathrm{K}}=\nu_{\mathrm{r}}$ at
$x=\mathcal{X}_{\mathrm{K}}=2$ for $b=b_{\mathrm{sr}}=1.5$.}
\end{figure*}

In the case of RN neutron stars with $b>1$ (naked singularity type
spacetimes), the situation is complicated, and the discussion can
be separated into four parts according to the parameter $b$ (see
figures \ref{orbity-vse:b} and \ref{ex-kep-rezy}):

\begin{enumerate}
    \item[(a)] $1 < b \leq 5/4$.

   The radial epicyclic frequency $\nu_{\mathrm{r}}$ has one local maximum as usual in RN spacetimes.

    \item[(b)] $5/4 < b < b_{\mathrm{m}}\doteq 1.313$.

    The radial epicyclic frequency $\nu_{\mathrm{r}}$ has one
    local maximum and one local minimum.

    \item[(c)] $b_{\mathrm{m}}\leq b \leq b_{\mathrm{c}}\doteq
    1.381$.

    The radial epicyclic frequency $\nu_{\mathrm{r}}$ has no local extrema (for all $b\geq b_{\mathrm{m}}$) since the condition $x>b$ has to be
    satisfied, and for these values of $b$ there is $\mathcal{X}_{\mathrm{r}}< x=b$.

    \item[(d)] $b > b_{\mathrm{c}}$.

    The maximum of the Keplerian frequency profile could be
    relevant, since it can appear outside the internal uniform
    density configuration given by $b_{\mathrm{C}}(x)$.

\end{enumerate}

We can summarize that in braneworld RN spacetimes:
\begin{itemize}
\item the Keplerian frequency is a monotonically decreasing
    function of radius for the whole range of the brany tidal charge parameter $b$
    in astrophysically relevant radii above
    the photon circular orbit (or the neutron star surface); an exception exists at spacetimes
    with $b > 1.381$, where the photon circular orbit is not
    defined.
\item the radial
    epicyclic frequency has a local maximum for RN black hole type spacetimes with $b \leq 1$, and
    vanishes at the innermost stable circular geodesic;
\item for RN naked singularity type spacetimes the behaviour of the radial
    frequency is different; a detailed analysis shows that the
radial epicyclic frequency
    can have one local maximum for $1<b \leq 5/4$, two local extrema for $5/4 < b < 1.313$ or no local extrema for $b\geq 1.313$.
\end{itemize}

\subsection{Orbital models of QPOs}

A number of low-mass X-ray binaries containing a neutron star show
quasiperiodic oscillations (QPOs) in their X-ray flux, i.e., peaks
in the Fourier variability power density spectra (PDS).
Frequencies of some QPOs are in the kHz range corresponding to
frequencies of the orbital motion close to the central neutron
star. Such so-called high-frequency (kHz) QPOs span large
frequency range of $200$ -- $1400$\,Hz. The two distinct modes
following their own correlation between frequency and properties
(amplitude and quality factor) of the peak are observed in this
range (see, e.g.,
\cite{Kli:2004:,Bel-Men-Hom:2005:ASTRA:,Bel-Men-Hom:2007:astro-ph/0702157:,Bar-Oli-Mil:2005:MONNR:,Bar-Oli-Mil:2005:AstrNachr:,Bar-Oli-Mil:2006:MNRAS:}).
 They are called lower and upper QPO because the frequencies of
upper QPO, $\nu_{\mathrm{U}}$, is higher than the frequency of the
lower QPO, $\nu_{\mathrm{L}}$, when both modes are detected
simultaneously; we call them twin peak QPOs in such situations
\cite{Tor-Stu-Bak:2007:CEURJP:,Tor-Bak-Stu-Cec:2007::prep,Tor-etal:2008:AA2:,STB:2007:}.
 It was shown that the twin peak QPOs are clustered around
frequencies corresponding to the frequency ratio of small integers
\cite{Tor-etal:2008:AA:,Tor-etal:2008:AA2:}. In some atoll
sources, namely 4U~1636$-$53 and 4U 1728, the frequency ratios are
clustered around $3\!:\!2$ and $5\!:\!4$, in others (4U 1608 and
4U 0614) they cluster around $3\!:\!2$ only, and in 4U 1820 and 4U
1735 they cluster around ratio $4\!:\!3$
\cite{Tor-etal:2007:RAGtime8and9CrossRef:MutRelkHzQPO}. In the
best known Z-source Sco X-1 the frequency ratios cluster around
$3\!:\!2$ and $5\!:\!4$ \cite{Abr-etal:ScoX1:2003:} and in the
exceptional Z-source Circinus X-1, the frequency ratio is
clustered around $3\!:\!1$ at relatively low frequencies $\sim
100\,\mathrm{Hz}$ \cite{Bou-etal:2006:}.

Several models have been outlined to explain observational data of
the neutron star kHz QPOs assuming that their origin is related to
orbital motion near the inner edge of accretion discs around the
neutron stars. The RP model of Stella and Vietri
\cite{Ste-Vie:1998:ASTRJ2L:} introduces the QPOs representing a
direct manifestation of modes of a relativistic epicyclic motion
of radiating blobs in the innermost parts of the accretion disc.
In this model, the upper and lower frequencies are identified in
the following way:
\begin{equation}
    \nu_{\mathrm{U}}=\nu_{\mathrm{K}},\qquad
    \nu_{\mathrm{L}}=\nu_{\mathrm{K}}-\nu_{\mathrm{r}}.
\end{equation}

The resonance model introduced by Abramowicz and Klu{\'z}niak
\cite{Abr-Klu:2001:ASTRA:,Klu-Abr:2001:ACTPB:} (and earlier in
other connections by Aliev and Galtsov \cite{Ali-Gal:1981:})
assumes a non-linear resonance (forced or parametric) of accretion
disc oscillations with vertical and radial epicyclic frequencies
$\nu_{\theta}$ and $\nu_{\mathrm{r}}$. However, in spherically
symmetric spacetimes assumed here $\nu_{\theta} =
\nu_{\mathrm{K}}$, and we identify
\begin{equation}
    \nu_{\mathrm{U}}=\nu_{\mathrm{K}},\qquad
    \nu_{\mathrm{L}}=\nu_{\mathrm{r}}.
\end{equation}

The observed frequency ratio $\nu_{\mathrm{U}}/\nu_{\mathrm{L}}$
decreases with increasing frequencies as $x\rightarrow
x_{\mathrm{ms}}$, but $\nu_{\mathrm{K}}/\nu_{\mathrm{r}}$
increases as $x\rightarrow x_{\mathrm{ms}}$ (see \cite{STB:2007:}
for details). On the other hand, frequency relations implied by
the RP model yield a trend, which is in good accord with
observation. In other words, frequency relation of the RP model
satisfies in a natural way the general tendency of twin peak kHz
QPOs as the ratio of
$\nu_{\mathrm{K}}\!:\!\left(\nu_{\mathrm{K}}-\nu_{\mathrm{r}}\right)$
decreases with increasing frequencies $\nu_{\mathrm{K}}$ and
$\nu_{\mathrm{K}}-\nu_{\mathrm{r}}$ for $x$ approaching the inner
edge of the disc at $x=x_{\mathrm{ms}}$, where
$\nu_{\mathrm{r}}\rightarrow 0$ and
$\nu_{\mathrm{K}}\!:\!\left(\nu_{\mathrm{K}}-\nu_{\mathrm{r}}\right)
\rightarrow 1$. The clustering of the observational data could
then be explained by non-linear resonant phenomena between the
Keplerian and periastron oscillatory motion in the framework of
the multiresonant model
\cite{STB:2007:,Stu-Kot-Tor:2007:RAGtime8and9CrossRef:MrmQPO}.
Note, however, that this frequency relation can coincide with
those implied under consideration of a forced resonance between
the epicyclic frequencies with observed higher eigenfrequency
$\nu_{\mathrm{U}}=\nu_{\mathrm{K}}=\nu_{\theta}$ and lower
combinational frequency
$\nu_{\mathrm{L}}=\nu_{\mathrm{K}}-\nu_{\mathrm{r}}$. In the
spherically symmetric spacetimes, they coincide with radial
$m=1$ and vertical $m=2$ disc oscillation modes. 
 Here, we focus for simplicity on the RP model.

\subsection{Resonance radii}

Radial profiles of the Keplerian and radial epicyclic frequency in
the external spacetimes of the black hole type ($b\leq 1$) enable
us to find the observed resonant points ($3\!:\!2$, $4\!:\!3$,
$5\!:\!4$) and relations for the resonant radii in the framework
of both the relativistic precession model frequency relations
$\nu_{\mathrm{K}}\!:\!\left(\nu_{\mathrm{K}}-\nu_{\mathrm{r}}\right)$
and the epicyclic frequency relations
$\nu_{\mathrm{K}}\!:\!\nu_{\mathrm{r}}$. The same relations hold
in the case of external spacetimes of the naked singularity type
with $1<b\leq 5/4$, where the behaviour of the Keplerian and
radial epicyclic frequencies is the same as in the black hole type
spacetimes.\footnote{We do not consider the internal branch of
stable orbits since it is too close to the limit given by
$b_{\mathrm{C}}$.}

\begin{figure*}[p]
\begin{center}
\subfigure[][]{\label{res-pol:a}\includegraphics[width=.86\hsize]{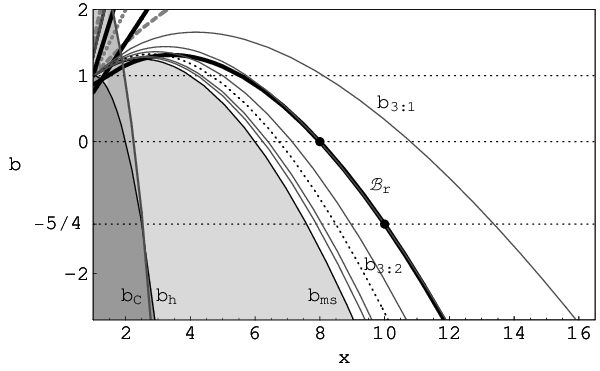}}\\
\subfigure[][]{\label{res-pol:b}\includegraphics[width=.86\hsize]{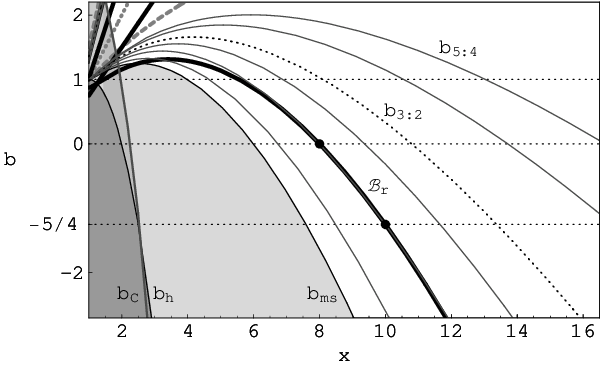}}
\end{center}
\caption{\label{res-pol}The resonant radii for ratios of small
integers for the RP model
$\nu_\mathrm{K}\!:\!(\nu_\mathrm{K}-\nu_\mathrm{r})$ (a) and for
the model of direct resonances $\nu_\mathrm{K}\!:\!\nu_\mathrm{r}$
(b). The grey lines represent the frequency ratios
$\nu_{\mathrm{U}}:\nu_{\mathrm{L}}=3\!:\!1$, $2\!:\!1$, $5\!:\!3$,
$3\!:\!2$ (dotted line), $4\!:\!3$ and $5\!:\!4$ subsequently
right to left in (a), left to right in (b). For $b=0$ and $b=-5/4$
the resonance radius where $\nu_{\mathrm{U}}:\nu_{\mathrm{L}} =
2\!:\!1$ coincides with the radius of the radial epicyclic
frequency local maximum (at $x=8$ for $b=0$, at $x=10$ for
$b=-5/4$).}
\end{figure*}

The resonant radii $x_{n:m}$, where the frequency ratio
$\nu_{\mathrm{K}}\!:\!\left(\nu_{\mathrm{K}}-\nu_{\mathrm{r}}\right)$
given by the RP model takes the value $n\!:\!m = p$, are
implicitly given by the condition
\begin{eqnarray}\label{res:pol:obecne}
 \fl b=b^{\mathrm{K}/(\mathrm{K}-\mathrm{r})}(x,p) \equiv
    b_{\mathrm{RP}}(x,p)\nonumber\\
    \equiv \frac{(p-1)^2 x^2+9p^2 x\mp x\sqrt{ x^2 (p-1)^4+2p^2 x\left(9p^2-2p+1\right)-15 p^4}}{8
    p^2}\,.
\end{eqnarray}
For the model of direct resonances of the epicyclic frequencies we
arrive at the relation
\begin{equation}
  b=b^{\mathrm{K}/\mathrm{r}}(x,p)\equiv
\frac{x}{8 p^2}  \left[9 p^2+x\mp\sqrt{x^2+2 p^2 x+p^4 (16
x-15)}\right]\,.
\end{equation}

These models are valid at $x> x_{\mathrm{ms}}$, but their
behaviour is inverse as $x\rightarrow x_{\mathrm{ms}}$, since
$\nu_{\mathrm{K}}\!:\!\left(\nu_{\mathrm{K}}-\nu_{\mathrm{r}}\right)\rightarrow
1$ there, while $\nu_{\mathrm{K}}\!:\!\nu_{\mathrm{r}}$ diverges
there since $\nu_{\mathrm{r}}(x=x_{\mathrm{ms}})=0$.

The resonant radii for ratios of small integers allowing for
efficient resonant phenomena \cite{Lan-Lif:1976:Mech:} are given
in \fref{res-pol}. Note that
$b^{\mathrm{K}/(\mathrm{K}-\mathrm{r})}_{3:2}(x)=b^{\mathrm{K}/\mathrm{r}}_{3:1}(x)$,
$b^{\mathrm{K}/(\mathrm{K}-\mathrm{r})}_{2:1}(x)=b^{\mathrm{K}/\mathrm{r}}_{2:1}(x)$,
$b^{\mathrm{K}/(\mathrm{K}-\mathrm{r})}_{3:1}(x)=b^{\mathrm{K}/\mathrm{r}}_{3:2}(x)$.
We stress that to model the data of twin peak QPOs in neutron star
systems, we can use the RP model, but the direct epicyclic
frequency model gives trends just opposite to what is observed.

Using the RP model, we can conclude that the resonant points are
located in the region between the marginally stable orbit and the
local extremum of the epicyclic frequency, i.e., in the innermost
part of the accretion disc. The frequency ratios $\sim 3\!:\!1$,
observed in the Z-source Circinus X-1 \cite{Bou-etal:2006:} could
still be explained in the framework of the RP model, but the
resonant points must be located above the maximum of the radial
epicyclic frequency. The fitting of the observational data will be
discussed in the following section, but for our rough estimates we
shall concentrate on the behaviour of the frequency-frequency
relations, postponing the detailed studies of the resonant
phenomena in individual sources to future studies.

\subsubsection{Strong resonance radii.}

Here, we would like to mention some special resonance conditions
that could be realized in the RN spacetimes of the naked
singularity type.

For RN naked singularity spacetime with $b>1.42$, the Keplerian
and radial epicyclic frequencies could be equal at a specific
radius (see \fref{ex-kep-rezy}). The condition
$\nu_{\mathrm{K}}(x,b)=\nu_{\mathrm{r}}(x,b)$ implies the relation
\begin{equation}
b=b_{\mathrm{sr}}(x)\equiv\frac{x}{8}\left(9+x\mp
\sqrt{x(x+18)-15}\right)
\end{equation}
determining in an implicit way the strong resonant radii, where
the resonance between the oscillations with Keplerian and radial
epicyclic frequencies could be strongest, and the frequency
scatter allowed for the resonance is largest. The function
$b_{\mathrm{sr}}(x)$ is illustrated in \fref{orbity-vse}. Note,
however, that the above relation is not relevant in case of
neutron star systems.

\section{\label{sec:pet}QPOs and RP model testing the braneworld models}

\begin{figure*}[p]
\begin{center}
\subfigure[][]{\label{fity:a}\includegraphics[width=.98\hsize]{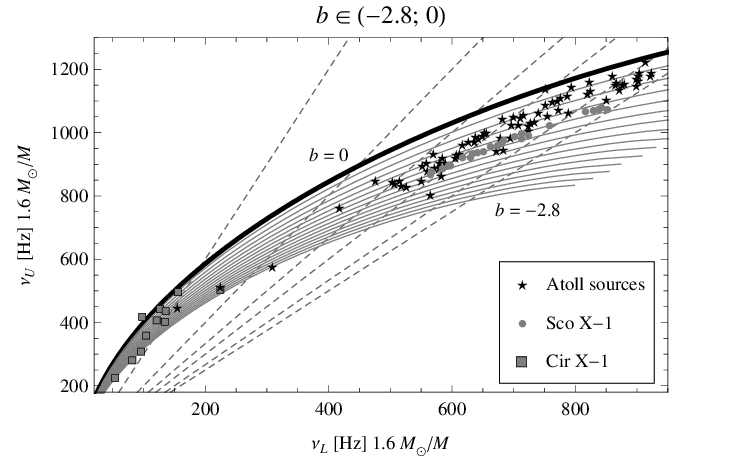}}\\
\subfigure[][]{\label{fity:b}\includegraphics[width=.98\hsize]{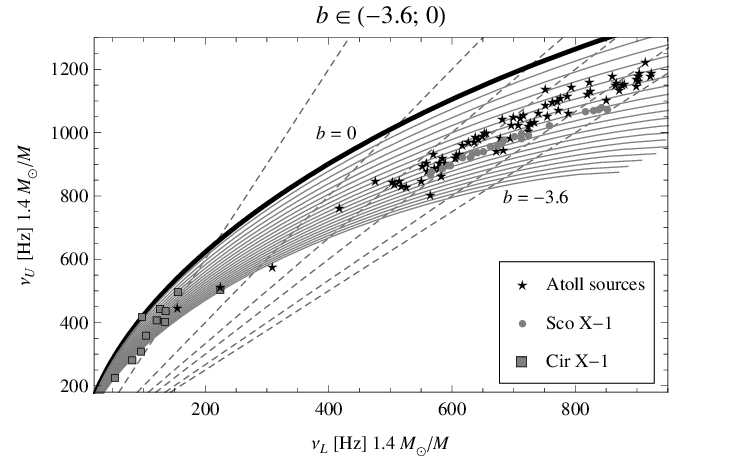}}
\end{center}
\caption{\label{fity}The relativistic precession model (with
negative tidal charge) fits for atoll sources, Sco X-1 and Cir X-1
for various values of the brany parameter $b$; the curves are
spaced by $0.2$ in $b$. The grey dashed lines represent the
frequency ratios $\nu_{\mathrm{U}}:\nu_{\mathrm{L}}=3\!:\!1$,
$2\!:\!1$, $5\!:\!3$, $3\!:\!2$, $4\!:\!3$ and $5\!:\!4$
subsequently left to right.}
\end{figure*}

We have summarized the twin peak QPOs data for the atoll sources
and Z-sources Sco X-1 and Circinus X-1 in \fref{fity} using the
data accumulated in \cite{Bel-Men-Hom:2007:astro-ph/0702157:}.
Clearly, there is a general tendency in the
$\nu_{\mathrm{U}}$--$\nu_{\mathrm{L}}$ frequency relations that
can be immediately deduced from the observational data:
$\nu_{\mathrm{U}}/\nu_{\mathrm{L}}$ decreases with
$\nu_{\mathrm{L}}$ ($\nu_{\mathrm{U}}$) increasing as stressed in
the previous section.

We shall use the most frequent RP model in order to obtain rough
restrictions on the allowed values of the tidal charge (and brane
tension) related to neutron star systems giving the observational
data demonstrated in \fref{fity}.

Assuming that the trend observed in the frequency relations
passing through the clusters of the neutron star twin peak kHz
QPOs is related to the modes of the RP model, we make two kinds of
fitting the data by the brany neutron star models. From these
fittings we deduce rough constraints to the tidal charge and brane
tension related to the neutron star systems generating the
observed data.

First, we assume negative values of the brany tidal charge (and
positive brane tension $\lambda > 0$) and consider any RP model
with values of $b<0$ touching any observational point. Thus, we
exclude the values of $b<0$ that correspond to the RP frequency
relations located outside the observational data (see
\fref{fity}). The results are characterized by
$M=1.6\,\mathrm{M}_{\odot}$ and $b\in(-2.8,0)$ (see
\fref{fity:a}). Assuming the canonical neutron star mass
$M=1.4\,\mathrm{M}_{\odot}$, we obtain the acceptable interval of
tidal charge $b\in (-3.6, 0)$ as demonstrated in \fref{fity:b}.

\begin{figure*}[t]
\begin{center}
\includegraphics[width=.98\hsize]{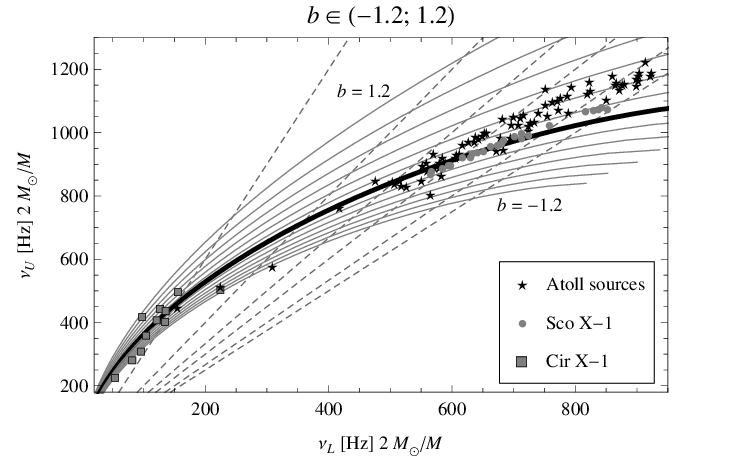}
\end{center}
\caption{\label{fity2}The relativistic precession model (with
positive and negative tidal charge) fits for the same ensemble of
data as in \fref{fity}. Note that the $M=2\,\mathrm{M}_{\odot}$,
$b=0$ RP frequency relation is commonly used for the fitting of
observed QPO data (see, e.g.,
\cite{Bel-Men-Hom:2007:astro-ph/0702157:}).}
\end{figure*}

Second, we allow both positive and negative values of $b$ (and
brane tension $\lambda$). Such a case we represent by an estimate
done in a symmetric way, i.e., we find such a neutron star mass
that the observational data are scanned for this mass parameter by
RP fit relations with brany tidal charge of the same size in both
positive and negative values; we have found
$M=2\,\mathrm{M}_{\odot}$ and $b\in(-1.2, +1.2)$ as demonstrated
in \fref{fity2}.\footnote{Note that the range of $b<5/4$ enables
us to limit our consideration to the RN naked singularity type
spacetimes with the same behaviour of the Keplerian and radial
epicyclic frequencies as in the black hole spacetimes; for
example, the strong resonance radii are irrelevant.}

The presented results are used to obtain estimates on restrictions
of the tidal charge $B$ using the relation
\begin{equation}
    |B| < |B_{(i)}|=|b_{(i)}|\left(\frac{\mathrm{G}M_{(i)}}{\mathrm{c}^2}\right)^2.
\end{equation}
We then obtain the following restrictions:
\begin{enumerate}
    \item[(a)] Assuming negative tidal charge
            with $|b_{(1)}|=2.8$ and $M_{(1)}=1.6\,\mathrm{M}_{\odot}$, we
            obtain
            \begin{equation}
                |B| < |B_{(1)}|=1.562\times
                10^{11}\,\mathrm{cm}^2,
            \end{equation}
            while the canonical mass $M_{(2)}=1.4\,\mathrm{M}_{\odot}$ and
            $|b_{(2)}|=3.6$ yield
            \begin{equation}
            |B| < |B_{(2)}|=1.537\times 10^{11}\,\mathrm{cm}^2.
            \end{equation}

    \item[(b)] Allowing both positive and negative values of the tidal
        charge, we obtain for $M_{(3)}=2\,\mathrm{M}_{\odot}$ and
        $|b_{(3)}|=1.2$ the estimate
        \begin{equation}
        |B| < |B_{(3)}|=1.046\times 10^{11}\,\mathrm{cm}^2
        \end{equation}
        that is naturally lower in comparison with the previous
        two cases because of the scatter of $b$ into both positive
        and negative values that reduces the magnitude of the
        dimensionless tidal charge by a factor of $\sim 1.5$.
\end{enumerate}

The brane tension can be estimated using the
relation~\eref{match:cond}, which yields
        \begin{equation}
        |\lambda| > |\lambda_{(i)}|=\frac{3\mathrm{G}M_{(i)}}{\mathrm{c}^2}\frac{R_{(i)} \rho_{(i)}}{B_{(i)}}
        \end{equation}
that can be transformed to the relation
        \begin{equation}
        |\lambda| > |\lambda_{(i)}|=\frac{3\rho_{(i)}}{|b_{(i)}|}\frac{R_{(i)}}{r_{\mathrm{G}}}=\frac{3\rho_{(i)}}{|b_{(i)}|}X_{(i)}.
        \end{equation}
Using the limit on compactness of the uniform configuration
$R/r_{\mathrm{G}}>R_{\mathrm{C}}/r_{\mathrm{G}}$, where
$R_{\mathrm{C}}(b)/r_{\mathrm{G}}=X_{\mathrm{C}}(b)$ is implicitly
given as a function of $b$ by relation~\eref{krit:pol}, we arrive
at the general estimate on the brane tension:
        \begin{equation}
        |\lambda| > |\lambda_{(i)}|=\frac{3\rho_{(i)}}{|b_{(i)}|}\frac{R_{\mathrm{C}}(b_{(i)})}{r_{\mathrm{G}}}=\frac{3\rho_{(i)}}{|b_{(i)}|}X_{\mathrm{C}}(b_{(i)}).
        \end{equation}
For the canonical values of the neutron stars, namely
$M=1.4\,\mathrm{M}_{\odot}$, $R=10\,\mathrm{km}$ 
and $\rho \sim 6.6\times 10^{14}\,\mathrm{g}/\mathrm{cm}^3$, we
find
        \begin{equation}\label{kanonicka}
        |\lambda| > |\lambda_{\mathrm{can}}|=2.7\times
        10^{15}\,\mathrm{g}/\mathrm{cm}^3.
        \end{equation}
In order to make a more precise restriction of $\lambda$, we have
to estimate the dimensionless surface radius $X$ of the
configuration for given $b_{(i)}$ and $M_{(i)}$. For
$b_{(i)}>-1.5$, the estimate should be related to
$b_{\mathrm{C}}(x)$, for $b_{(i)}<-1.5$ we have to use
$b_{\mathrm{h}}(x)$.

We find
        \begin{equation}
        |\lambda| >
        |\lambda_{(i)}|=\frac{3}{|b_{(i)}|}\left(\frac{\mathrm{M}_{\odot}}{M_{(i)}}\right)^2\left(\frac{1}{X_{(i)}}\right)^2
        1.477\times
        10^{17}\,\mathrm{g}/\mathrm{cm}^3.
        \end{equation}
For $M_{(1)}=1.6\,\mathrm{M}_{\odot}$, $|b_{(1)}|=2.8$ and
choosing $X_{(1)}=4$, we find
        \begin{equation}
        |\lambda_{1}|=3.9\times
        10^{15}\,\mathrm{g}/\mathrm{cm}^3,
        \end{equation}
which is comparable to our simple canonical
estimate~\eref{kanonicka}. For $M_{(3)}=2\,\mathrm{M}_{\odot}$,
$|b_{(3)}|=1.2$ and choosing $X_{(3)}=3$, we find
        \begin{equation}
        |\lambda_{3}|=1\times
        10^{16}\,\mathrm{g}/\mathrm{cm}^3.
        \end{equation}
This estimate exceeds the canonical one by a factor of $\sim 3.7$
because of the lowering of the dimensionless tidal charge limit.

\section{\label{sec:sest}Conclusions}

Using the RP model of kHz QPOs observed in neutron star X-ray
binaries modified by the hypothetical tidal charge of the neutron
stars as implied by the braneworld uniform star model, we are able
to put rough restriction on the tidal charge magnitude and brane
tension magnitude related to the considered group of atoll sources
and some Z-sources. This is the first attempt at an estimate
coming from the consideration of effects in the extremely strong
gravitational fields.

The presented method is rough but robust, since the effects of
neutron star rotation are not important. According to
\eref{alfa:r}, the effect of the tidal charge is given by the term
$\sim b/r^2$. On the other hand, the leading term giving
rotational effects is $\sim a/r^{3/2}$ and, moreover, the limits
on the rotational parameters taken from observed rotational
frequencies of neutron stars \cite{Tor-Bak-Stu-Cec:2007::prep} put
a strong limit of $a\leq 0.3$, so that its influence onto the
character of the fitting frequency-frequency curves is small as
shown by T{\"{o}}r{\"{o}}k et
al.~\cite{Tor-Bak-Stu-Cec:2007::prep}. Of course, fitting the
observational data of individual sources by precise methods will
need the rotational parameter for fine tuning.\footnote{The
precise estimates of individual sources probably require also
inclusion of the non-geodesic effects caused by the neutron star
magnetic field \cite{Bak-etal:2008:magn:}.}

Assuming negative values of $B$ and positive values of $\lambda$,
we obtained rough estimates $|B| < 1.5\times
10^{11}\,\mathrm{cm}^2$ and $|\lambda|> 3\times
10^{15}\,\mathrm{g}/\mathrm{cm}^3$ for both $M\sim
1.6\,\mathrm{M}_{\odot}$ and the canonical case of neutron stars
with $M\sim 1.4\,\mathrm{M}_{\odot}$. On the other hand, from the
limits allowing both positive and negative values of $b$
($\lambda$), starting from the standard mass estimate $M\sim
2\,\mathrm{M}_{\odot}$ related to QPOs
\cite{Bel-Men-Hom:2005:ASTRA:}, we obtained a lower estimate of
$|B| < 10^{11}\,\mathrm{cm}^2$, but higher values of $|\lambda|>
10^{16}\,\mathrm{g}/\mathrm{cm}^3$.

These values should be confronted with estimates given by
B{\"o}hmer et al. \cite{Boh-etal:2008:SolarSys:}, concerning the
effects of tidal charge in the limit of weak gravitational field
related to the solar system. They considered the perihelion
precession (Mercury), deflection of light near the edge of Sun,
and the radar echo delay experiment. Clearly, the most convenient
for comparison seems to be the first phenomenon, being of the same
nature as the epicyclic motion of blobs in the relativistic
precession model. The weak field restrictions are
\cite{Boh-etal:2008:SolarSys:}
        \begin{equation}
        |B| <5\times 10^{8}\,\mathrm{cm}^2,\qquad \lambda > 7\times
        10^{13}\,\mathrm{g}/\mathrm{cm}^3.
        \end{equation}
Note that the rough strong field regime model gives estimates of
the tidal charge higher than the weak field regime in more than
two orders of magnitude. However, as stressed in
\cite{Boh-etal:2008:SolarSys:}, a substantial part of the
perihelion shift could correspond to the effect of solar
oblateness, modifying thus, in principle, very strongly the
estimates of $B$ and $\lambda$. On the other hand, the
restrictions implied in \cite{Boh-etal:2008:SolarSys:} by the
light deflection and radar echo delay are $ |B| <
10^{19}\,\mathrm{cm}^2$ and $|\lambda |>
10^{17}\,\mathrm{g}/\mathrm{cm}^3$, which are substantially, by
orders, higher than those given by rough estimates of QPOs
observational data in X-ray binaries. We can thus conclude that
the restrictions on $|B|$ and $|\lambda |$ implied by rough
fitting of the neutron star X-ray binaries kHz QPOs data, i.e., in
the strong gravitational field regime, give better fits as
compared to those implied by the weak-field limit, if these
results are not masked by the effects of the solar oblateness.
Clearly, further and more detailed studies of all these phenomena
are necessary. Moreover, there is a variety of other vacuum
solutions of spherically symmetric static field on the brane
\cite{Sha-Dad:2004:,Cas-Maz:2003:,Cas-Fab-Maz:2002:} that deserve
attention and have to be tested separately since the Weyl tensor
in the complete 5D theory is unknown yet.

We plan to make further, more sophisticated comparisons of the
plausible kHz QPOs models for neutron star X-ray binaries,
including the RP model and the resonance disc oscillation ones,
and compare these models with data obtained for individual
sources. Precise fitting of these individual source data to the
models including the influence of the tidal charge could, in
principle, bring some new interesting results and limits on the
tidal charge and brane tension although we cannot expect
substantial shift of the rough estimates presented here, but
rather better fits of observational data and more realistic
estimates of the neutron star mass, more close to the expected
canonical value. In fact, generally, we can conclude that the
presence of a negative tidal charge enables lowering of the
neutron star mass as can be seen from our rough estimates.

\ack
This work was supported by the Czech grant MSM~4781305903.


\section*{References}

\begin{thebibliography}{10}

\bibitem{Ark-Dim-Dva:1998:}
N.~Arkani-Hamed, S.~Dimopoulos, and G.~Dvali.
\newblock {The Hierarchy Problem and New Dimensions at a Millimeter}.
\newblock {\em Phys. Lett. B}, 429:263--272, 1998.
\newblock {\tt arXiv:hep-ph/9803315v1}.

\bibitem{Ran-Sun:1999:}
L.~Randall and R.~Sundrum.
\newblock {An Alternative to Compactification}.
\newblock {\em Phys. Rev. Lett.}, 83(23):4690--4693, 1999.
\newblock {\tt arXiv:hep-th/9906064v1}.

\bibitem{Shi-Mae-Sas:2000:}
T.~Shiromizu, K.~i.~Maeda, and M.~Sasaki.
\newblock {The Einstein Equations on the 3-Brane World}.
\newblock {\em Phys. Rev. D}, 62:024012, 2000.
\newblock {\tt arXiv:gr-qc/9910076v3}.

\bibitem{Dad-etal:2000:}
N.~Dadhich, R.~Maartens, P.~Papadopoulos, and V.~Rezania.
\newblock {Black holes on the brane}.
\newblock {\em Phys. Lett. B}, 487:1, 2000.
\newblock {\tt arXiv:hep-th/0003061v3}.

\bibitem{Ger-Maa:2001:}
C.~Germani and R.~Maartens.
\newblock {Stars in the braneworld}.
\newblock {\em Phys. Rev. D}, 64:124010, 2001.
\newblock {\tt arXiv:hep-th/0107011v3}.

\bibitem{Maa:2004:}
R.~Maartens.
\newblock {Brane-world gravity}.
\newblock {\em Living Rev. Rel.}, 7:7, 2004.
\newblock {\tt arXiv:gr-qc/0312059v2}.

\bibitem{Mis-Tho-Whe:1973:Gra:}
C.~W. Misner, K.~S. Thorne, and J.~A. Wheeler.
\newblock {\em Gravitation}.
\newblock W. H. Freeman and Co, New York, San Francisco, 1973.

\bibitem{Sas-Shi-Mae:2000:}
M.~Sasaki, T.~Shiromizu, and K.~i.~Maeda.
\newblock {Gravity, stability, and energy conservation on the Randall--Sundrum brane world}.
\newblock {\em Phys. Rev. D}, 62:024008, 2000.
\newblock {\tt arXiv:hep-th/9912233v3}.

\bibitem{Ali-Gum:2005:}
A.~N. Aliev and A.~E. {G{\"u}mr{\"u}k{\c c}{\"u}o{\u g}lu}.
\newblock {Charged rotating black holes on a 3-brane}.
\newblock {\em Phys. Rev. D}, 71(10):104027, 2005.
\newblock {\tt arXiv:hep-th/0502223v2}.

\bibitem{Web-Gle:1992:ASTRJ2:upr:}
F.~Weber and N.~K. Glendenning.
\newblock {Application of the improved Hartle method for the construction of general relativistic rotating neutron star models}.
\newblock {\em Astrophys. J.}, 390:541--549, 1992.

\bibitem{Gle:1997:CompactStars:}
N.~K. Glendenning.
\newblock {\em {Compact Stars: Nuclear Physics, Particle Physics, and General Relativity}}.
\newblock Springer-Verlag, New York, 1997.

\bibitem{Boh-etal:2008:SolarSys:}
C.~G. B{\"o}hmer, T.~Harko, and F.~S.~N. Lobo.
\newblock {Solar system tests of brane world models}.
\newblock {\em Classical Quantum Gravity}, 25(4):5015, 2008.
\newblock {\tt arXiv:0801.1375v2 [gr-qc]}.

\bibitem{PdeLeon:2008:}
J.~Ponce de~Leon.
\newblock {Static exteriors for nonstatic braneworld stars}.
\newblock {\em Classical Quantum Gravity}, 25:5012, 2008.
\newblock {\tt arXiv:0711.4415v2 [gr-qc]}.

\bibitem{Kee-Pet:2006:brane-lensing:}
C.~R. Keeton and A.~O. Petters.
\newblock {Formalism for testing theories of gravity using lensing by compact objects. III. Braneworld gravity}.
\newblock {\em Phys. Rev. D}, 73(10):104032, 2006.
\newblock {\tt arXiv:gr-qc/0603061v3}.

\bibitem{Boz:2005:}
V.~Bozza.
\newblock {Gravitational lensing in the strong field limit}.
\newblock {\em Phys. Rev. D}, 66(10):103001, 2002.

\bibitem{Eir:2005:}
E.~F. Eiroa.
\newblock {Braneworld black hole gravitational lens: Strong field limit analysis}.
\newblock {\em Phys. Rev. D}, 71(8):083010, 2005.
\newblock {\tt arXiv:gr-qc/0410128v2}.

\bibitem{Whi:2005:}
R.~Whisker.
\newblock {Strong gravitational lensing by braneworld black holes}.
\newblock {\em Phys. Rev. D}, 71(6):064004, 2005.
\newblock {\tt arXiv:astro-ph/0411786v3}.

\bibitem{Sch-Stu:2007:RAGtime8and9CrossRef:SpBranKBH}
J.~Schee and Z.~Stuchl{\'{\i}}k.
\newblock {Spectral line profile of radiating ring orbiting a brany Kerr black hole}.
\newblock In Hled{\'{\i}}k and Stuchl{\'{\i}}k
  \cite{Hle-Stu:2007:RAGtime8and9:Proceedings}, pages 209--220.

\bibitem{Sch-Stu:2007:RAGtime8and9CrossRef:OEBraK}
J.~Schee and Z.~Stuchl{\'{\i}}k.
\newblock {Optical effects in brany Kerr spacetimes}.
\newblock In Hled{\'{\i}}k and Stuchl{\'{\i}}k
  \cite{Hle-Stu:2007:RAGtime8and9:Proceedings}, pages 221--256.

\bibitem{Rem-McCli:2006:ARASTRA:}
R.~A. Remillard and J.~E. McClintock.
\newblock {X-Ray Properties of Black-Hole Binaries}.
\newblock {\em Annual Review of Astronomy and Astrophysics}, 44(1):49--92, September 2006.
\newblock {\tt arXiv:astro-ph/0606352v1}.

\bibitem{Kli:2004:}
M.~van~der Klis.
\newblock {Rapid X-ray Variability}.
\newblock In W.~H.~G. Lewin and M.~van~der Klis, editors, {\em {Compact Stellar X-Ray Sources}}, pages 39--112, Cambridge, 2006. Cambridge University Press.

\bibitem{Kli:2000:ARASTRA:}
M.~van~der Klis.
\newblock {Millisecond Oscillations in X-ray Binaries}.
\newblock {\em Annual Review of Astronomy and Astrophysics}, 38:717--760, 2000.
\newblock {\tt arXiv:astro-ph/0001167v1}.

\bibitem{Ter-Abr-Klu:2005:ASTRA:QPOresmodel}
G.~T{\"{o}}r{\"{o}}k, M.~A. Abramowicz, W.~Klu{\'z}niak, and Z.~Stuchl{\'{\i}}k.
\newblock {The orbital resonance model for twin peak kHz quasi periodic oscillations in microquasars}.
\newblock {\em Astronomy and Astrophysics}, 436(1):1--8, 2005.

\bibitem{Bel-Men-Hom:2005:ASTRA:}
T.~Belloni, M.~M{\'e}ndez, and J.~Homan.
\newblock {The distribution of kHz QPO frequencies in bright low mass X-ray binaries}.
\newblock {\em Astronomy and Astrophysics}, 437:209--216, 2005.
\newblock {\tt arXiv:astro-ph/0501186v2}.

\bibitem{Bel-Men-Hom:2007:astro-ph/0702157:}
T.~Belloni, M.~M{\'e}ndez, and J.~Homan.
\newblock {On the kHz QPO frequency correlations in bright neutron star X-ray binaries}.
\newblock {\em Monthly Notices Roy. Astronom. Soc.}, 376:1133--1138, 2007.
\newblock {\tt arXiv:astro-ph/0702157v1}.

\bibitem{Bar-Oli-Mil:2005:MONNR:}
D.~Barret, J.-F. Olive, and M.~Coleman Miller.
\newblock {An abrupt drop in the coherence of the lower kilohertz QPO in 4U~1636$-$536}.
\newblock {\em Monthly Notices Roy. Astronom. Soc.}, 361:855--860, 2005.
\newblock {\tt arXiv:astro-ph/0505402v1}.

\bibitem{Tor:2007:ASTRA:RevTwPk6}
G.~T{\"{o}}r{\"{o}}k.
\newblock {Reverse of twin peak QPO amplitude relationship in six atoll sources}.
\newblock {\em Astronomy and Astrophysics}, submitted, 2008.

\bibitem{Str:2007:astro-ph/0701390:}
T.~E. Strohmayer, R.F. Mushotzky, L.~Winter, R.~Soria, P.~Uttley, and M.~Cropper.
\newblock {Quasi-periodic Variability in NGC 5408 X-1}.
\newblock {\em Astrophys. J.}, 660:580--586, 2007.
\newblock {\tt arXiv:astro-ph/0701390v1}.

\bibitem{Lac-Cze-Abr:2006:astro-ph0607594:}
P.~Lachowicz, B.~Czerny, and M.~A. Abramowicz.
\newblock {Wavelet analysis of MCG-6-30-15 and NGC~4051: a possible discovery of QPOs in $2\!:\!1$ and $3\!:\!2$ resonance}.
\newblock {\em Monthly Notices Roy. Astronom. Soc.}, submitted, 2006.
\newblock {\tt arXiv:astro-ph/0607594v1}.

\bibitem{Asc-etal:2004:ASTRA:}
B.~Aschenbach, N.~Grosso, D.~Porquet, and P.~Predehl.
\newblock {X-ray flares reveal mass and angular momentum of the Galactic Center black hole}.
\newblock {\em Astronomy and Astrophysics}, 417:71--78, 2004.
\newblock {\tt arXiv:astro-ph/0401589v2}.

\bibitem{Ter:2005:astro-ph0412500:}
G.~T{\"{o}}r{\"{o}}k.
\newblock {A possible $3\!:\!2$ orbital epicyclic resonance in QPOs frequencies of Sgr\,A$^*$}.
\newblock {\em Astronomy and Astrophysics}, 440(1):1--4, 2005.
\newblock {\tt arXiv:astro-ph/0412500v1}.

\bibitem{Tor-etal:2008:AA:}
G.~T{\"{o}}r{\"{o}}k, M.~A. Abramowicz, P.~Bakala, M.~Bursa, J.~Hor{\'{a}}k, P.~Rebusco, and Z.~Stuchl{\'{\i}}k.
\newblock {On the origin of clustering of frequency ratios in the atoll source 4U~1636$-$53}.
\newblock {\em Acta Astronom.}, 58:113--119, 2008.
\newblock {\tt arXiv:0802.4026v2 [astro-ph]}.

\bibitem{Tor-etal:2008:AA2:}
G.~T{\"{o}}r{\"{o}}k, M.~A. Abramowicz, P.~Bakala, M.~Bursa, J.~Hor{\'{a}}k, W.~Klu{\'z}niak, P.~Rebusco, and Z.~Stuchl{\'{\i}}k.
\newblock {Distribution of kilohertz QPO frequencies and their ratios in the atoll source 4U~1636$-$53}.
\newblock {\em Acta Astronom.}, 58:15--21, 2008.
\newblock {\tt arXiv:0802.4070v2 [astro-ph]}.

\bibitem{SK:2008:PhysRevD:}
Z.~Stuchl{\'{\i}}k and A.~Kotrlov{\'{a}}.
\newblock {Orbital resonances in discs around braneworld Kerr black holes}.
\newblock {\em General Relativity and Gravitation}, accepted, 2008.

\bibitem{Ste-Vie:1998:ASTRJ2L:}
L.~Stella and M.~Vietri.
\newblock {Lense--Thirring Precession and Quasi-periodic Oscillations in Low-Mass X-Ray Binaries}.
\newblock {\em Astrophys. J. Lett.}, 492:L59--L62, 1998.
\newblock {\tt arXiv:astro-ph/9709085v1}.

\bibitem{Tor-Bak-Stu-Cec:2007::prep}
G.~T{\"{o}}r{\"{o}}k, P.~Bakala, Z.~Stuchl{\'{\i}}k, and P.~{\v{C}}ech.
\newblock {Modelling the twin peak QPO distribution in the atoll source 4U~1636$-$53}.
\newblock {\em Acta Astronom.}, 58:1--14, 2008.

\bibitem{Tor-etal:2007:RAGtime8and9CrossRef:MutRelkHzQPO}
G.~T{\"{o}}r{\"{o}}k, M.~Bursa, J.~Hor{\'{a}}k, Z.~Stuchl{\'{\i}}k, and P.~Bakala.
\newblock {On mutual relation of kHz~QPOs}.
\newblock In Hled{\'{\i}}k and Stuchl{\'{\i}}k
  \cite{Hle-Stu:2007:RAGtime8and9:Proceedings}, pages 501--510.

\bibitem{Kat-Fuk-Min:1998:BHAccDis:}
S.~Kato, J.~Fukue, and S.~Mineshige.
\newblock {Black-hole accretion disks}.
\newblock In Shoji Kato, Jun Fukue, and Sin Mineshige, editors, {\em
  {Black-hole accretion disks}}, Kyoto, Japan, 1998. Kyoto University Press.

\bibitem{Klu-Abr:2000:ASTROPH:}
W.~Klu\'{z}niak and M.~A. Abramowicz.
\newblock {The physics of kHz QPOs~-- strong gravity's coupled anharmonic oscillators}, 2001.
\newblock {\tt arXiv:astro-ph/0105057v1}.

\bibitem{Rez-etal:2003:MONNR:}
L.~Rezzolla, S'i. Yoshida, T.~J. Maccarone, and O.~Zanotti.
\newblock {A new simple model for high-frequency quasi-periodic oscillations in black hole candidates}.
\newblock {\em Monthly Notices Roy. Astronom. Soc.}, 344(3):L37--L41, 2003.
\newblock {\tt arXiv:astro-ph/0307487v1}.

\bibitem{Tor-Stu:2005:RAGtime6and7:CrossRef}
G.~T{\"{o}}r{\"{o}}k and Z.~Stuchl{\'{\i}}k.
\newblock {Epicyclic frequencies of Keplerian motion in Kerr spacetimes}.
\newblock In Hled{\'{\i}}k and Stuchl{\'{\i}}k
  \cite{Hle-Stu:2005:RAGtime6and7:Proceedings}, pages 315--338.

\bibitem{Ter-Stu:2005:ASTRA:}
G.~T{\"{o}}r{\"{o}}k and Z.~Stuchl{\'{\i}}k.
\newblock {Radial and vertical epicyclic frequencies of Keplerian motion in the field of Kerr naked singularities. Comparison with the black hole case and possible instability of naked singularity accretion discs}.
\newblock {\em Astronomy and Astrophysics}, 437(3):775--788, 2005.
\newblock {\tt arXiv:astro-ph/0502127v1}.

\bibitem{Stu-Kot-Tor:2007:RAGtime8and9CrossRef:MrmQPO}
Z.~Stuchl{\'{\i}}k, A.~Kotrlov{\'{a}}, and G.~T{\"{o}}r{\"{o}}k.
\newblock {Multi-resonance models of QPOs}.
\newblock In Hled{\'{\i}}k and Stuchl{\'{\i}}k
  \cite{Hle-Stu:2007:RAGtime8and9:Proceedings}, pages 363--416.

\bibitem{Sra:2005:ASTRN:}
E.~{\v S}r{\'{a}}mkov{\'{a}}.
\newblock {Epicyclic oscillation modes of a Newtonian, non-slender torus}.
\newblock {\em Astronom. Nachr.}, 326(9):835--837, 2005.

\bibitem{Bla-etal:2006:ASTRJ2:}
O.~M. Blaes, E.~{\v S}r{\'{a}}mkov{\'{a}}, M.~A. Abramowicz, W.~Klu{\'z}niak, and U.~Torkelsson.
\newblock {Epicyclic Oscillations of Fluid Bodies: Newtonian Nonslender Torus}.
\newblock {\em Astrophys. J.}, 665:642--653, 2007.
\newblock {\tt arXiv:0706.4483v1 [astro-ph]}.

\bibitem{Lan-Lif:1976:Mech:}
L.~D. Landau and E.~M. Lifshitz.
\newblock {\em Mechanics}, volume~I of {\em Course of Theoretical Physics}.
\newblock Elsevier Butterworth-Heinemann, Oxford, 3rd edition, 1976.

\bibitem{Nay-Moo:1979:NonOscilations:}
A.~H. Nayfeh and D.~T. Mook.
\newblock {\em {Nonlinear Oscillations}}.
\newblock Wiley-interscience, New York, 1979.

\bibitem{Abr-etal:2005:RAGtime6and7:CrossRef}
M.~A. Abramowicz, D.~Barret, M.~Bursa, J.~Hor{\'{a}}k, W.~Klu{\'z}niak, P.~Rebusco, and G.~T{\"{o}}r{\"{o}}k.
\newblock {A note on the slope-shift anticorrelation in the neutron star kHz QPOs data}.
\newblock In Hled{\'{\i}}k and Stuchl{\'{\i}}k
  \cite{Hle-Stu:2005:RAGtime6and7:Proceedings}, pages 1--9.

\bibitem{Abr-etal:2005:ASTRN:}
M.~A. Abramowicz, D.~Barret, M.~Bursa, J.~Hor{\'{a}}k, W.~Klu{\'z}niak, P.~Rebusco, and G.~T{\"{o}}r{\"{o}}k.
\newblock {The correlations and anticorrelations in QPO data}.
\newblock {\em Astronom. Nachr.}, 326(9):864--866, 2005.
\newblock {\tt arXiv:astro-ph/0510462v1}.

\bibitem{Ali-Gal:1981:}
A.~N. Aliev and D.~V. Galtsov.
\newblock {Radiation from relativistic particles in nongeodesic motion in a strong gravitational field}.
\newblock {\em General Relativity and Gravitation}, 13:899--912, 1981.

\bibitem{Jar-Abr-Pac:1980:ACTAS:}
M.~Jaroszy{\'n}ski, M.~A. Abramowicz, and B.~Paczy{\'n}ski.
\newblock {Supercritical accretion disks around black holes}.
\newblock {\em Acta Astronom.}, 30:1--34, 1980.

\bibitem{Pac-Wii:1980:ASTRA:}
B.~Paczy{\'n}ski and P.~J. Wiita.
\newblock {Thick accretion disks and supercritical luminosities}.
\newblock {\em Astronomy and Astrophysics}, 88(1--2):23--31, 1980.

\bibitem{Koz-Jar-Abr:1978:ASTRA:}
M.~Koz{\l}owski, M.~Jaroszy{\'n}ski, and M.~A. Abramowicz.
\newblock {The analytic theory of fluid disks orbiting the Kerr black hole}.
\newblock {\em Astronomy and Astrophysics}, 63(1--2):209--220, 1978.

\bibitem{Kro-Haw:2002:ASTRJ2:}
J.~H. Krolik and J.~F. Hawley.
\newblock {Where Is the Inner Edge of an Accretion Disk around a Black Hole?}
\newblock {\em Astrophys. J.}, 573(2):754--763, 2002.
\newblock {\tt arXiv:astro-ph/0203289v1}.

\bibitem{Bar-Oli-Mil:2005:AstrNachr:}
D.~Barret, J.-F. Olive, and M.~C. Miller.
\newblock {Drop of coherence of the lower kilo-Hz QPO in neutron stars: Is there a link with the innermost stable circular orbit?}
\newblock {\em Astronom. Nachr.}, 326:808--811, 2005.
\newblock {\tt arXiv:astro-ph/0510094v1}.

\bibitem{Bar-Oli-Mil:2006:MNRAS:}
D.~Barret, J.-F. Olive, and M.~C. Miller.
\newblock {The coherence of kilohertz quasi-periodic oscillations in the X-rays from accreting neutron stars}.
\newblock {\em Monthly Notices Roy. Astronom. Soc.}, 370:1140--1146, 2006.
\newblock {\tt arXiv:astro-ph/0605486v1}.

\bibitem{Tor-Stu-Bak:2007:CEURJP:}
G.~T{\"{o}}r{\"{o}}k, Z.~Stuchl{\'{\i}}k, and P.~Bakala.
\newblock {A remark about possible unity of the neutron star and black hole high frequency QPOs}.
\newblock {\em Central European J. Phys.}, 5(4):457--462, 2007.

\bibitem{STB:2007:}
Z.~Stuchl{\'{\i}}k, G.~T{\"{o}}r{\"{o}}k, and P.~Bakala.
\newblock {On a multi-resonant origin of high frequency quasiperiodic oscillations in the neutron-star X-ray binary 4U~1636$-$53}.
\newblock {\em Astronomy and Astrophysics}, submitted, 2007.
\newblock {\tt arXiv:0704.2318v2 [astro-ph]}.

\bibitem{Abr-etal:ScoX1:2003:}
M.~A. Abramowicz, T.~Bulik, M.~Bursa, and W.~Klu{\'z}niak.
\newblock {Evidence for a $2\!:\!3$ resonance in Sco X-1 kHz QPOs}.
\newblock {\em Astronomy and Astrophysics}, 404:L21--L24, 2003.
\newblock {\tt arXiv:astro-ph/0206490v2}.

\bibitem{Bou-etal:2006:}
S.~Boutloukos, M.~van~der Klis, D.~Altamirano, M.~Klein-Wolt, R.~Wijnands, P.~G. Jonker, and R.~P. Fender.
\newblock {Discovery of Twin kHz QPOs in the Peculiar X-Ray Binary Circinus X-1}.
\newblock {\em Astrophys. J.}, 653:1435--1444, 2006.
\newblock {\tt arXiv:astro-ph/0608089v2}.

\bibitem{Abr-Klu:2001:ASTRA:}
M.~A. Abramowicz and W.~Klu{\'z}niak.
\newblock {A precise determination of black hole spin in GRO~J1655$-$40}.
\newblock {\em Astronomy and Astrophysics}, 374:L19, 2001.
\newblock {\tt arXiv:astro-ph/0105077v1}.

\bibitem{Klu-Abr:2001:ACTPB:}
W.~Klu{\'z}niak and M.~A. Abramowicz.
\newblock {Strong-field gravity and orbital resonance in black holes and neutron stars~-- kHz quasi-periodic oscillations (QPO)}.
\newblock {\em Acta Phys. Polon. B}, 32:3605--3612, 2001.

\bibitem{Bak-etal:2008:magn:}
P.~Bakala, E.~{\v S}r{\'{a}}mkov{\'{a}}, Z.~Stuchl{\'{\i}}k, and G.~T{\"{o}}r{\"{o}}k.
\newblock {On magnetic-field induced non-geodesic corrections to the relativistic precession QPO model}.
\newblock {\em Astrophys. J.}, submitted, 2008.

\bibitem{Sha-Dad:2004:}
S.~Shankaranarayanan and N.~Dadhich.
\newblock {Non-Singular Black-Holes on the Brane}.
\newblock {\em Int. J. Modern Phys. D}, 13:1095--1103, 2004.
\newblock {\tt arXiv:gr-qc/0306111v2}.

\bibitem{Cas-Maz:2003:}
R.~Casadio and L.~Mazzacurati.
\newblock {Bulk Shape of Brane-World Black Holes}.
\newblock {\em Modern Phys. Lett. A}, 18:651--660, 2003.
\newblock {\tt arXiv:gr-qc/0205129v2}.

\bibitem{Cas-Fab-Maz:2002:}
R.~Casadio, A.~Fabbri, and L.~Mazzacurati.
\newblock {New black holes in the brane world?}
\newblock {\em Phys. Rev. D}, 65(8):084040, 2002.
\newblock {\tt arXiv:gr-qc/0111072v2}.

\bibitem{Hle-Stu:2007:RAGtime8and9:Proceedings}
S.~Hled{\'{\i}}k and Z.~Stuchl{\'{\i}}k, editors.
\newblock {\em Proceedings of RAGtime 8/9: Workshops on black holes and neutron
  stars, Opava, Hradec nad Moravic\'{i}, 15--19/19--21 September 2006/2007},
  Opava, 2007. Silesian University in Opava.

\bibitem{Hle-Stu:2005:RAGtime6and7:Proceedings}
S.~Hled{\'{\i}}k and Z.~Stuchl{\'{\i}}k, editors.
\newblock {\em Proceedings of RAGtime 6/7: Workshops on black holes and neutron
  stars, Opava, 16--18/18--20 September 2004/2005}, Opava, 2005. Silesian
  University in Opava.

\end{thebibliography}
\end{document}